\documentclass{emulateapj}
\usepackage{multirow}
\usepackage{color}
\usepackage[normalem]{ulem}
\usepackage{amssymb}

\begin{document}

\title{GLOBAL FAR-ULTRAVIOLET PROPERTIES OF THE CYGNUS LOOP}

\author{Il-Joong Kim\altaffilmark{1}}
\author{Kwang-Il Seon\altaffilmark{1}}
\author{Yeo-Myeong Lim\altaffilmark{2}}
\author{Dae-Hee Lee\altaffilmark{1}}
\author{Wonyong Han\altaffilmark{1}}
\author{Kyoung-Wook Min\altaffilmark{2}}
\author{Jerry Edelstein\altaffilmark{3}}

\altaffiltext{1}{Korea Astronomy and Space Science Institute, Daejeon 305-348, Republic of Korea; ijkim@kasi.re.kr}
\altaffiltext{2}{Korea Advanced Institute of Science and Technology, Daejeon 305-701, Republic of Korea}
\altaffiltext{3}{Space Sciences Laboratory, University of California, Berkeley, CA 94702, USA}

\begin{abstract}
We present the \ion{C}{3} $\lambda$977, \ion{O}{6} $\lambda\lambda$1032, 1038 and \ion{N}{4}{]} $\lambda$1486 emission line maps of the Cygnus Loop, obtained with the newly processed data of Spectroscopy of Plasma Evolution from Astrophysical Radiation (SPEAR; also known as FIMS) mission. In addition, the \ion{Si}{4}+\ion{O}{4}{]} line complexes around 1400 \AA{} are resolved into two separate emission lines, whose intensity demonstrates a relatively high \ion{Si}{4} region predicted in the previous study. The morphological similarity between the \ion{O}{6} and X-ray images, as well as a comparison of the \ion{O}{6} intensity with the value expected from the X-ray results, indicates that large portions of the observed \ion{O}{6} emissions could be produced from X-ray emitting gas. Comparisons of the far-ultraviolet (FUV) images with the optical and \ion{H}{1} 21 cm images, reveal spatial variations of shock-velocity populations and high FUV extinction in the direction of a previously identified \ion{H}{1} cloud. By calculating the FUV line ratios for several subregions of the Cygnus Loop, we investigate the spatial variation of the population of radiative shock velocities; and the effects of resonance scattering, X-ray emitting gas, and non-radiative shocks. The FUV and X-ray luminosity comparisons between the Cygnus Loop and the Vela supernova remnant suggest that the fraction of shocks in the early evolutionary stages is much larger in the Cygnus Loop.
\end{abstract}


\section{INTRODUCTION}

The Cygnus Loop is one of the brightest supernova remnants (SNRs) in various wavelength domains due to its proximity of $\sim$540 pc \citep{blair05} and large angular size ($\sim$3$\arcdeg$ in diameter). It is considered to be a middle-aged SNR ($\sim$1.4$\times$10$^{4}$ yr; Levenson et al. 1998), and to be interacting actively with the ambient medium. These points make the remnant one of the most extensively observed and studied SNRs. Many X-ray studies were performed to understand the spatial structures and spectral properties of the remnant \citep{aschenbach99,charles85,katsuda08,katsuda11,ku84,leahy04,levenson97,levenson99,levenson02,levenson05,miyata98,tsunemi07,uchida08}. Also, \citet{katagiri11} provided results from gamma-ray measurements. Details of many optical filaments, including Balmer filaments, were analyzed in various optical studies \citep{blair99,blair05,levenson01,raymond80b,shull91}. \citet{levenson98} presented the optical H$\alpha$, [\ion{O}{3}], and [\ion{S}{2}] images of the entire region. The global IR images were presented by \citet{arendt92} and \citet{braun86}, and the high resolution IR images of a non-radiative shock were presented by \citet{sankrit10}. Radio continuum studies at various frequencies provided results for spectral indices and polarization properties \citep{green90,leahy97,sun06,uyaniker02}. Also, \citet{leahy02,leahy03,leahy05} presented the results of the \ion{H}{1} 21 cm observations over various velocity channels. In all wavelength ranges, the global structure of the Cygnus Loop is well defined to be a shell-like feature with a south extension, and the outer rim of the shell is much brighter than the inner region. This morphology can be explained by a scenario in which the Cygnus Loop exploded in a cavity, and its blast wave is now hitting the walls of the cavity \citep{levenson98,levenson99}.

Many of the far-ultraviolet (FUV) observations for various optical filaments were made and several FUV emission lines originating from non-radiative or radiative shocks were detected and analyzed \citep{blair91b,blair02,cornett92,danforth00,long92,raymond80a,raymond81,raymond83,raymond88,sankrit00,sankrit02,sankrit07}. However, the entire region of the Cygnus Loop was analyzed in only a few FUV studies. \citet{blair91b} reported maps of whole regions in the \ion{C}{3} $\lambda$977 and \ion{O}{6} $\lambda\lambda$1032, 1038 emission lines using the {\it Voyager 2} Ultraviolet Spectrometer, and \citet{rasmussen92} presented those in the \ion{O}{6} $\lambda\lambda$1032, 1038 emission lines by a rocket-borne experiment, High Resolution Emission Line Spectrometer (HIRES). The Spectroscopy of Plasma Evolution from Astrophysical Radiation (SPEAR), also known as Far-Ultraviolet Imaging Spectrograph (FIMS), carried out the global FUV observation of the Cygnus Loop in two wavelength channels, 900--1150 \AA{} and 1340--1750 \AA{}. \citet{seon06} reported the early results using only the long wavelength channel data. They showed the \ion{Si}{4}+\ion{O}{4}{]} $\lambda\lambda$1400, 1403 (unresolved), \ion{C}{4} $\lambda\lambda$1548, 1551, \ion{He}{2} $\lambda$1640.5, and \ion{O}{3}{]} $\lambda\lambda$1661, 1666 emission-line images, which were similar to the X-ray, optical, and radio-continuum images. They also presented the spectra in the 1340--1750 \AA{} range and estimated shock velocities from the line ratio of \ion{O}{4}{]} to \ion{O}{3}{]}.

In this paper, we analyze the short wavelength channel data of the SPEAR/FIMS observations and obtain new spectral images in the long wavelength channel. We create the \ion{C}{3} $\lambda$977 and \ion{O}{6} $\lambda\lambda$1032, 1038 emission-line images of the Cygnus Loop and compare them with the previous results. We separate out the blended \ion{Si}{4}+\ion{O}{4}{]} image, originally presented in \citet{seon06}, into two individual images of \ion{Si}{4} and \ion{O}{4}{]}. A new image at \ion{N}{4}{]} $\lambda$1486 is also presented with the newly processed SPEAR/FIMS data. Next, we compare the resulting FUV images with the X-ray, optical, and \ion{H}{1} 21 cm images, and discuss the FUV properties of the Cygnus Loop on a global scale. From the newly processed data, we update the previous results for the FUV line intensities, ratios, and luminosities.

\section{DATA REDUCTION}

SPEAR/FIMS is a dual-channel FUV imaging spectrograph for observing diffuse FUV emissions from the interstellar medium (ISM). The short bandpass channel (S-channel) covers 900--1150 \AA{} with 4$\arcdeg$.0$\times$4$\arcmin$.6 field of view, and the long bandpass channel (L-channel) covers 1340--1750 \AA{} with 7$\arcdeg$.4$\times$4$\arcmin$.3 field of view. Their spectral and spatial resolutions are $\lambda/\Delta\lambda\sim550$ and 5$\arcmin$, respectively. \citet{edelstein06b} described the SPEAR/FIMS mission and its science objectives in detail. The SPEAR/FIMS instrument, its on-orbit performance, and the basic processing of the data are described in \citet{edelstein06a}.

By combining the sky-survey data with the target-pointed data used in \citet{seon06}, we utilized the data obtained from a total of 43 orbits. We improved the accuracy of the attitude knowledge up to 5$\arcmin$ by using an updated version (for sky-survey data) of the software-correction code used in \citet{seon06}. The procedure of correction was briefly described in \citet{seon06}. We used the HEALPix pixelization scheme \citep{gorski05} to make images and spectra. A pixel size of $\sim$3$\arcmin$.4 (HEALPix resolution parameter \texttt{Nside} = 1024) and a wavelength bin size of 0.5 \AA{} were selected for the S-channel. On the other hand, \texttt{Nside} = 2048 and 1 \AA{} were used for the L-channel. To mask the pixels including FUV-bright stars, we identified early-type stars in the area of present concern using the {\it TD-1} catalog, and removed a total of 15 stars in the L-channel and 3 stars in the S-channel.

\section{DATA ANALYSIS AND RESULTS}

For emission-line images, the spectrum extracted from each pixel was divided into small segments including individual emission lines. Then each segment was fitted with Gaussian line profiles plus a linear continuum. For better statistics, the pixel size was increased to $\sim$6$\arcmin$.8 (\texttt{Nside} = 512) and the pixels with exposure time of $<$1 second were excluded before the spectral fitting. We fixed the central wavelengths and widths of the Gaussian functions, to be the calibrated line centers and the spectral resolutions, respectively. The intrinsic line ratios of 2:1 for the \ion{O}{6} $\lambda\lambda$1031.9, 1037.6 and \ion{C}{4} $\lambda\lambda$1548.2, 1550.8 doublet lines approach 1:1 in an optically thick case due to resonant scattering \citep{long92}. Therefore, we set the ratios for the doublet lines to vary between 1:1 and 2:1. The \ion{O}{3}{]} lines at 1660.8 and 1666.1 \AA{} were assumed to have an intensity ratio of 0.34:1, according to their statistical weights. We also attempted to fit the \ion{Si}{4}+\ion{O}{4}{]} line complexes as follows. We assumed that the \ion{Si}{4} lines are a doublet line at 1393.8 and 1402.8 \AA{}, while the \ion{O}{4}{]} lines are a quintet of lines at 1397.2, 1399.8, 1401.2, 1404.8, and 1407.4 \AA{}. However, since the \ion{Si}{4} $\lambda$1402.8 \AA{} line was unable to be resolved from the \ion{O}{4}{]} quintet lines, only the \ion{Si}{4} $\lambda$1393.8 \AA{} line was fitted along with the \ion{O}{4}{]} quintet lines. The strength of the \ion{Si}{4} $\lambda$1402.8 line was assumed to be half that of the \ion{Si}{4} $\lambda$1393.8 line, although it would be valid only in an optically thin case. Then, the strength of the \ion{Si}{4} $\lambda$1402.8 line was subtracted from the value obtained for the \ion{O}{4}{]} quintet lines and added into that of the \ion{Si}{4} doublet.

\begin{figure}[t]
\begin{centering}
\includegraphics[scale=0.25]{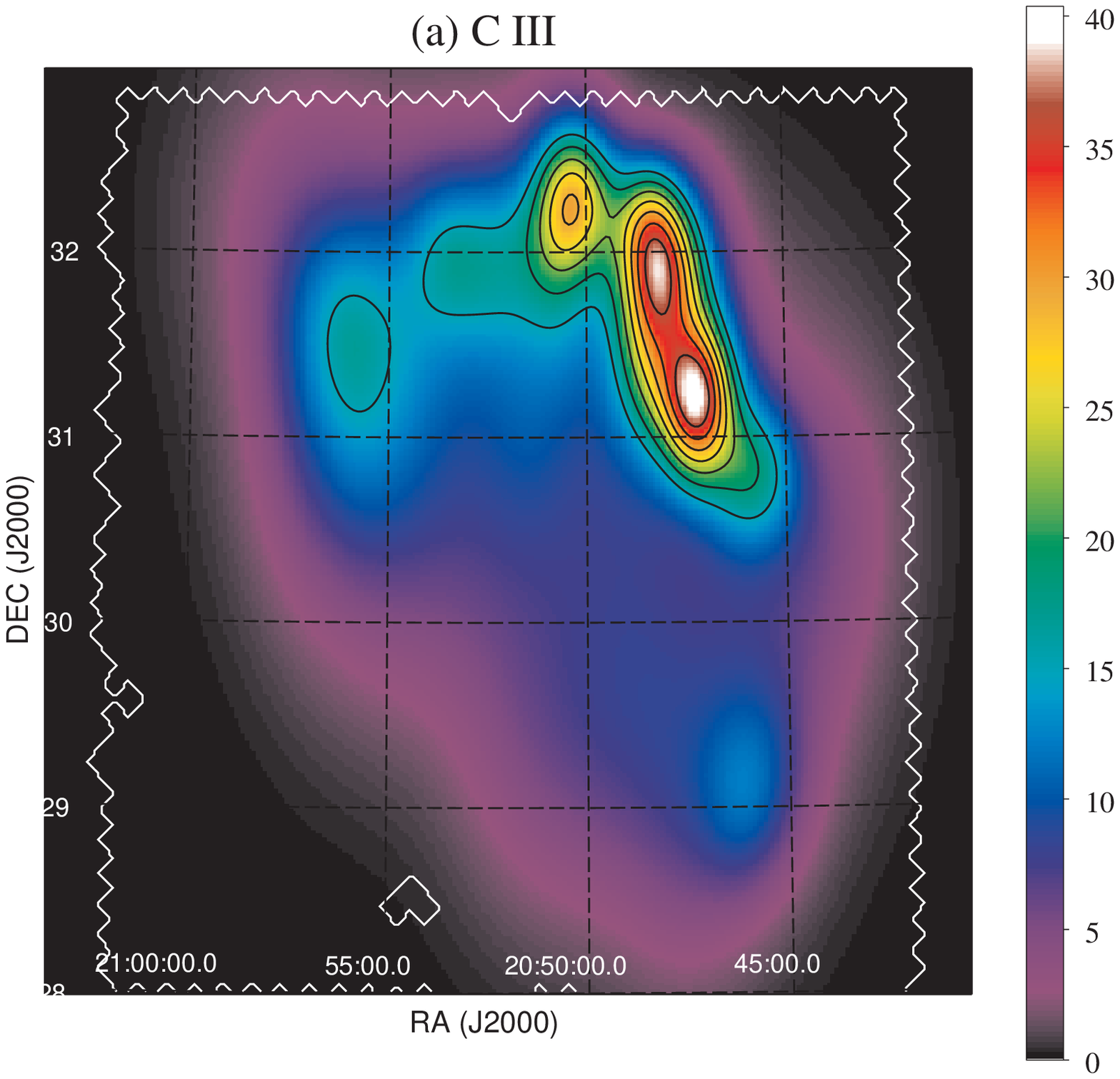}\quad\includegraphics[scale=0.25]{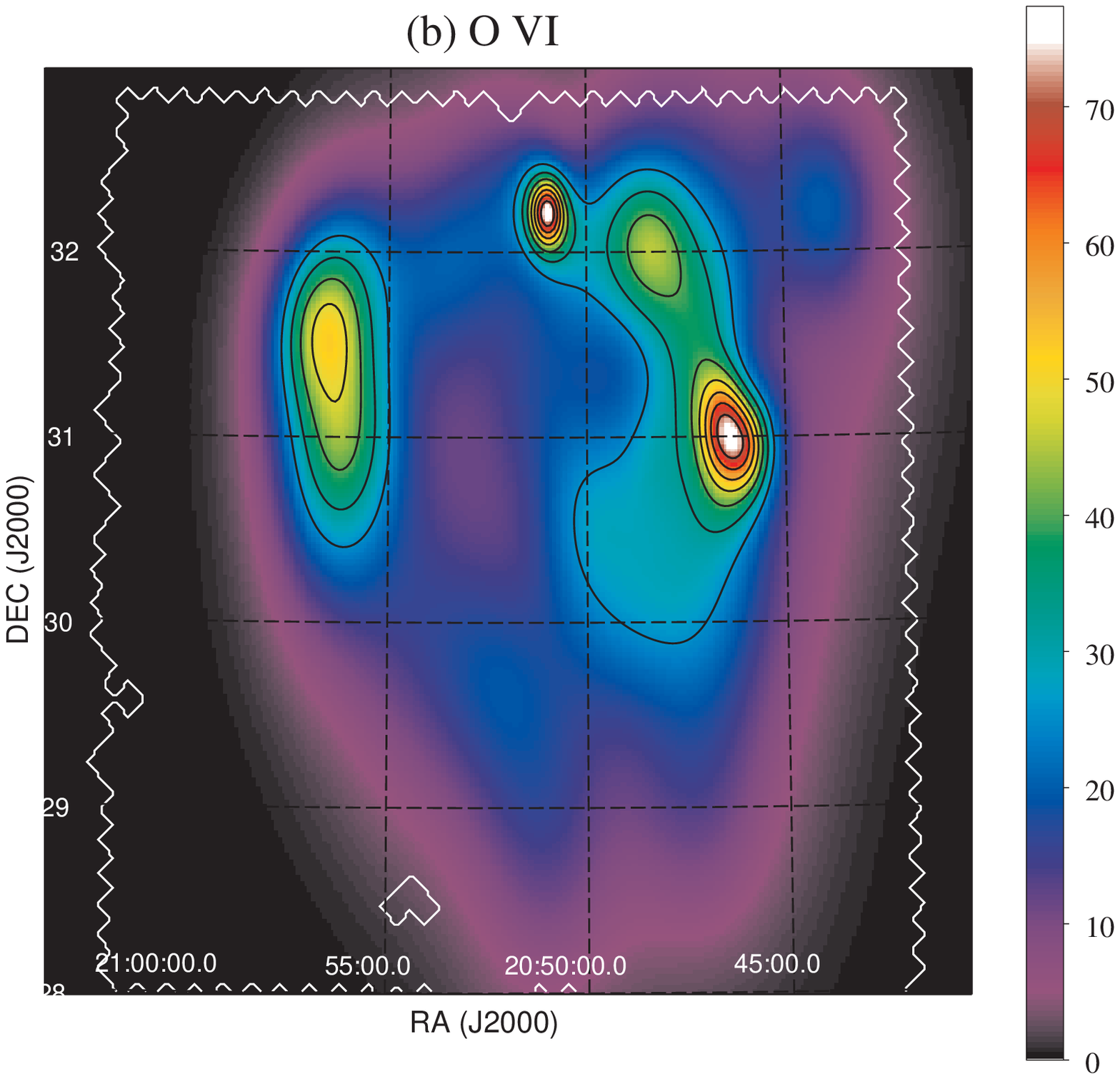}
\par\end{centering}
\vspace{0.2cm}
\begin{centering}
\includegraphics[scale=0.25]{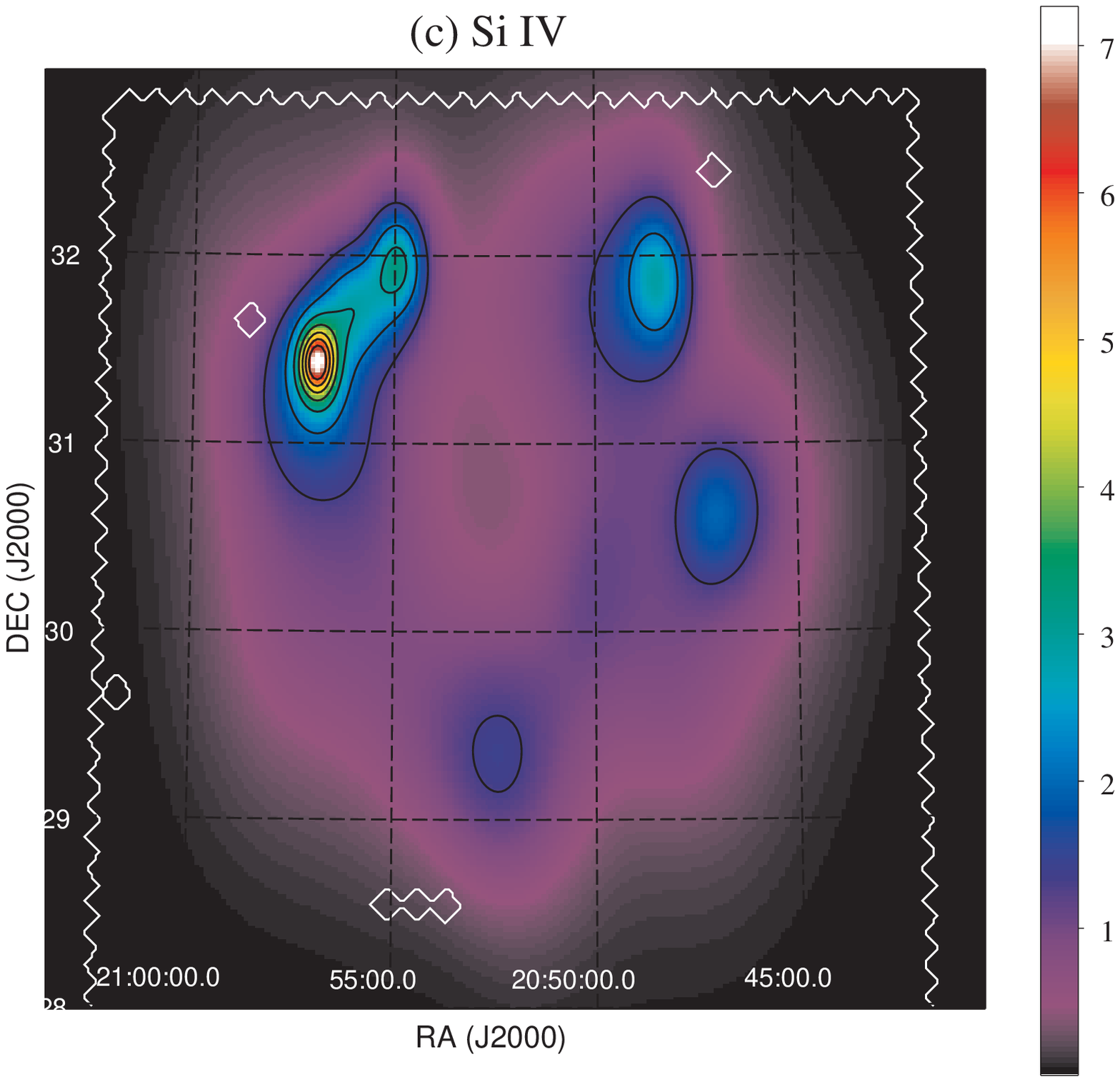}\quad\includegraphics[scale=0.25]{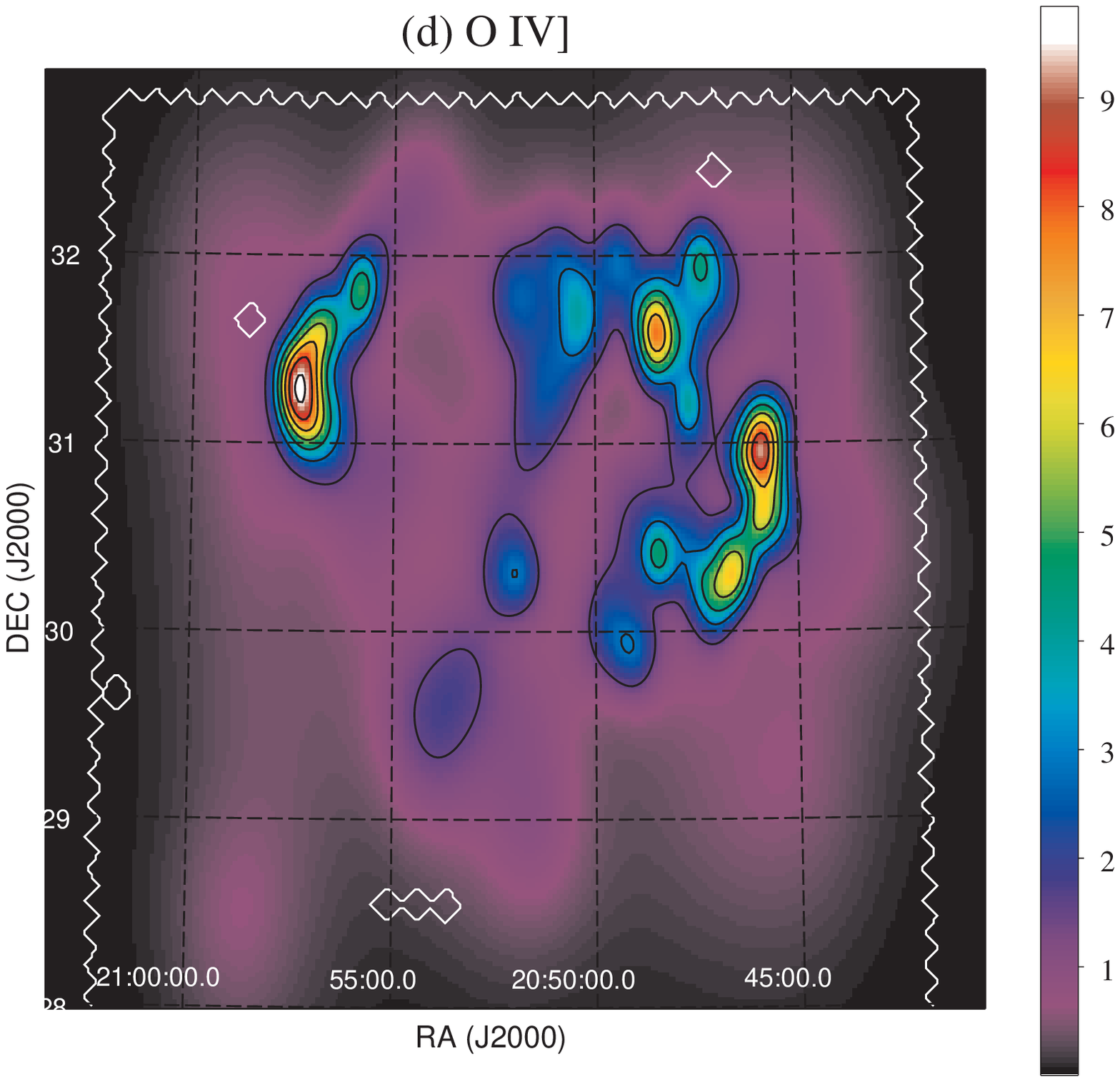}
\par\end{centering}
\vspace{0.2cm}
\begin{centering}
\includegraphics[scale=0.25]{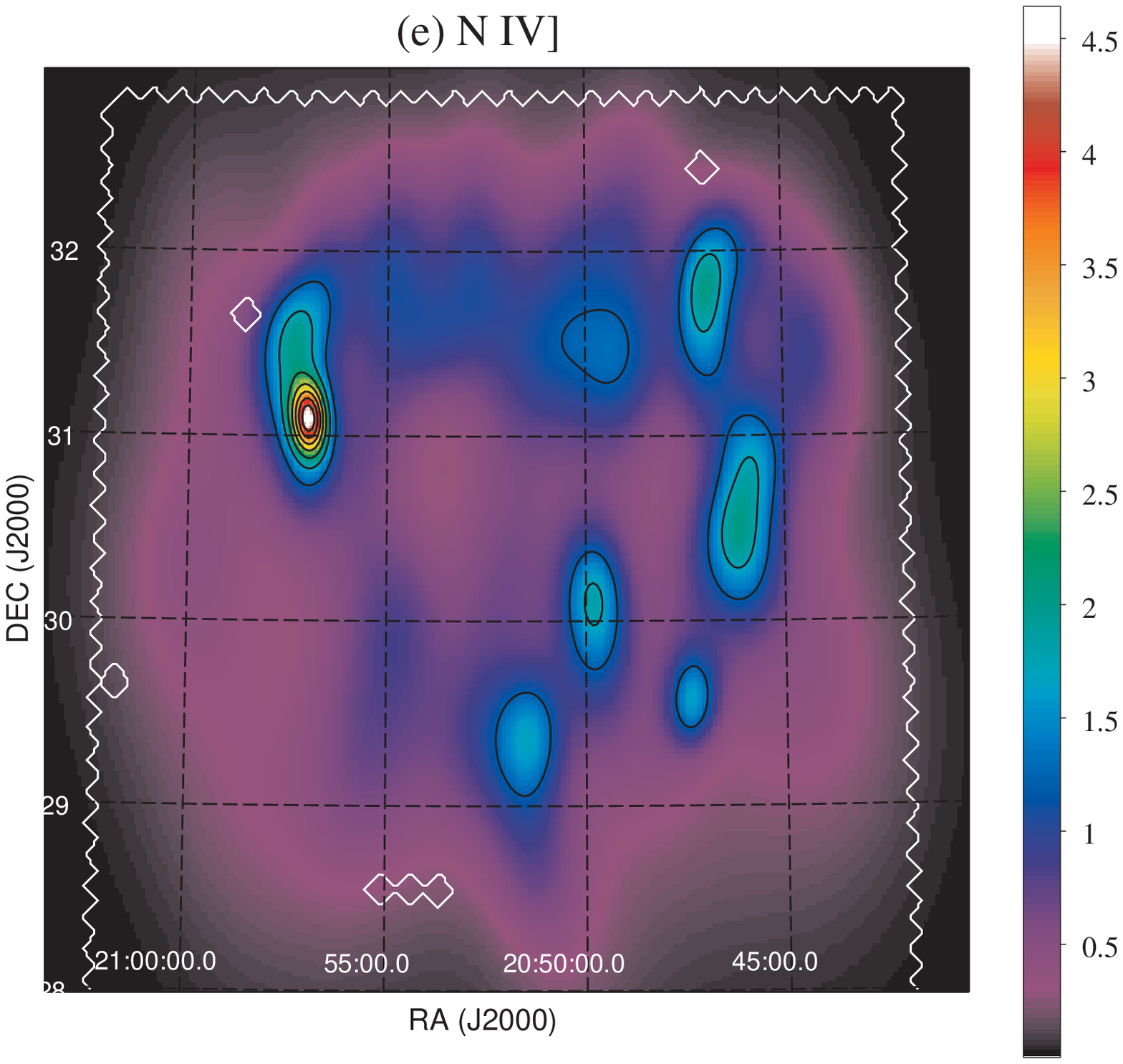}\quad\includegraphics[scale=0.25]{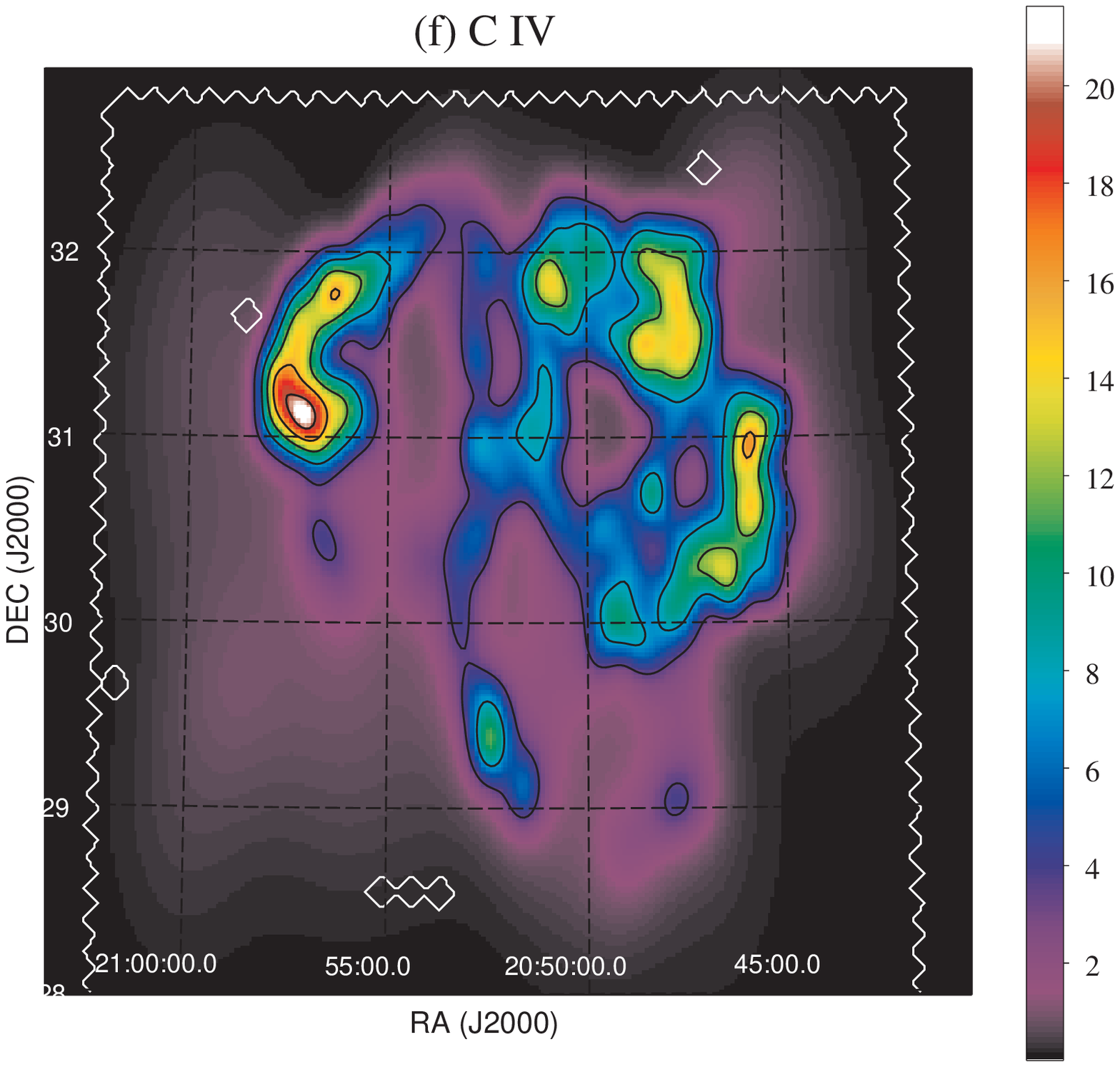}
\par\end{centering}
\vspace{0.2cm}
\begin{centering}
\includegraphics[scale=0.25]{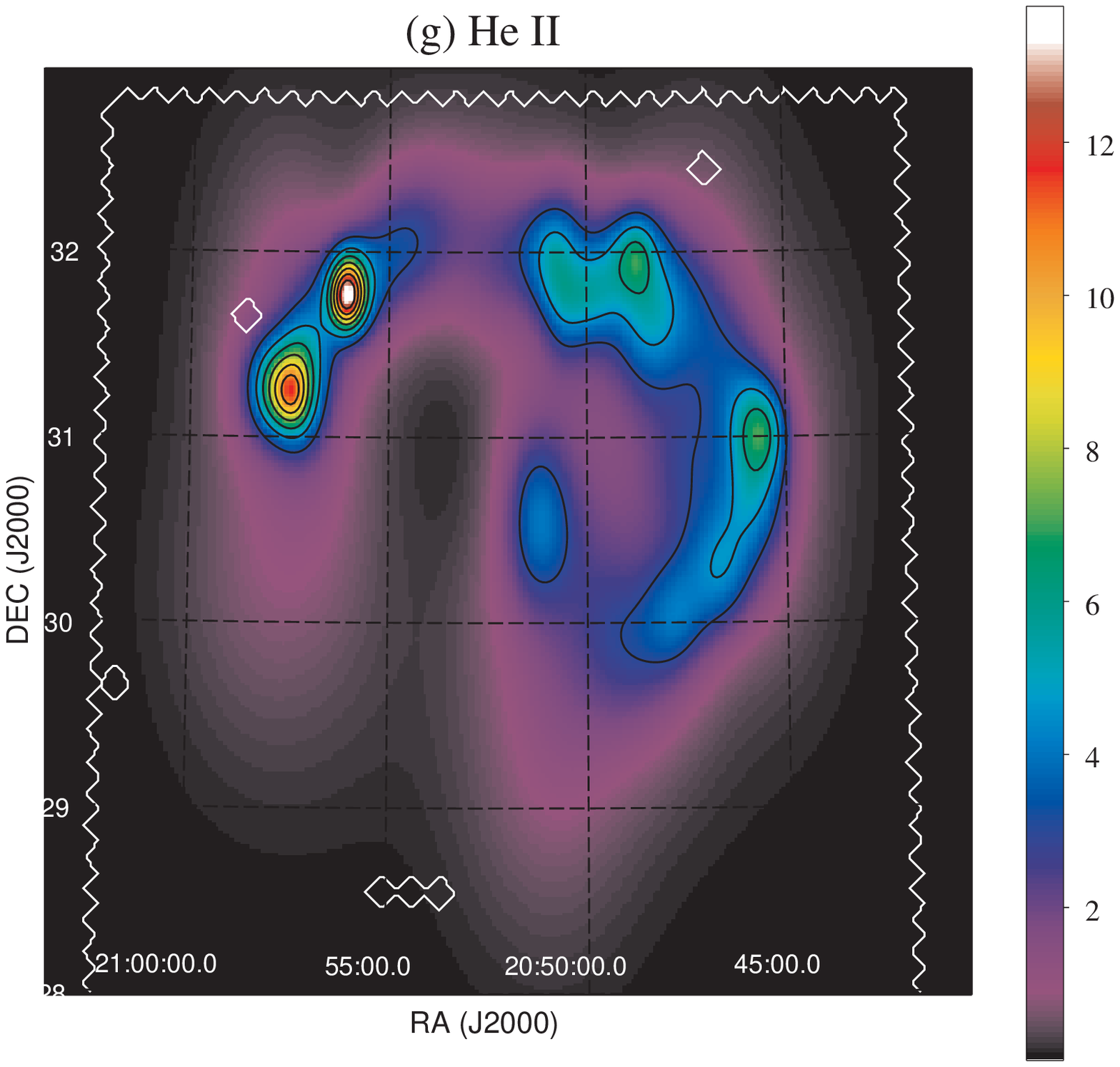}\quad\includegraphics[scale=0.25]{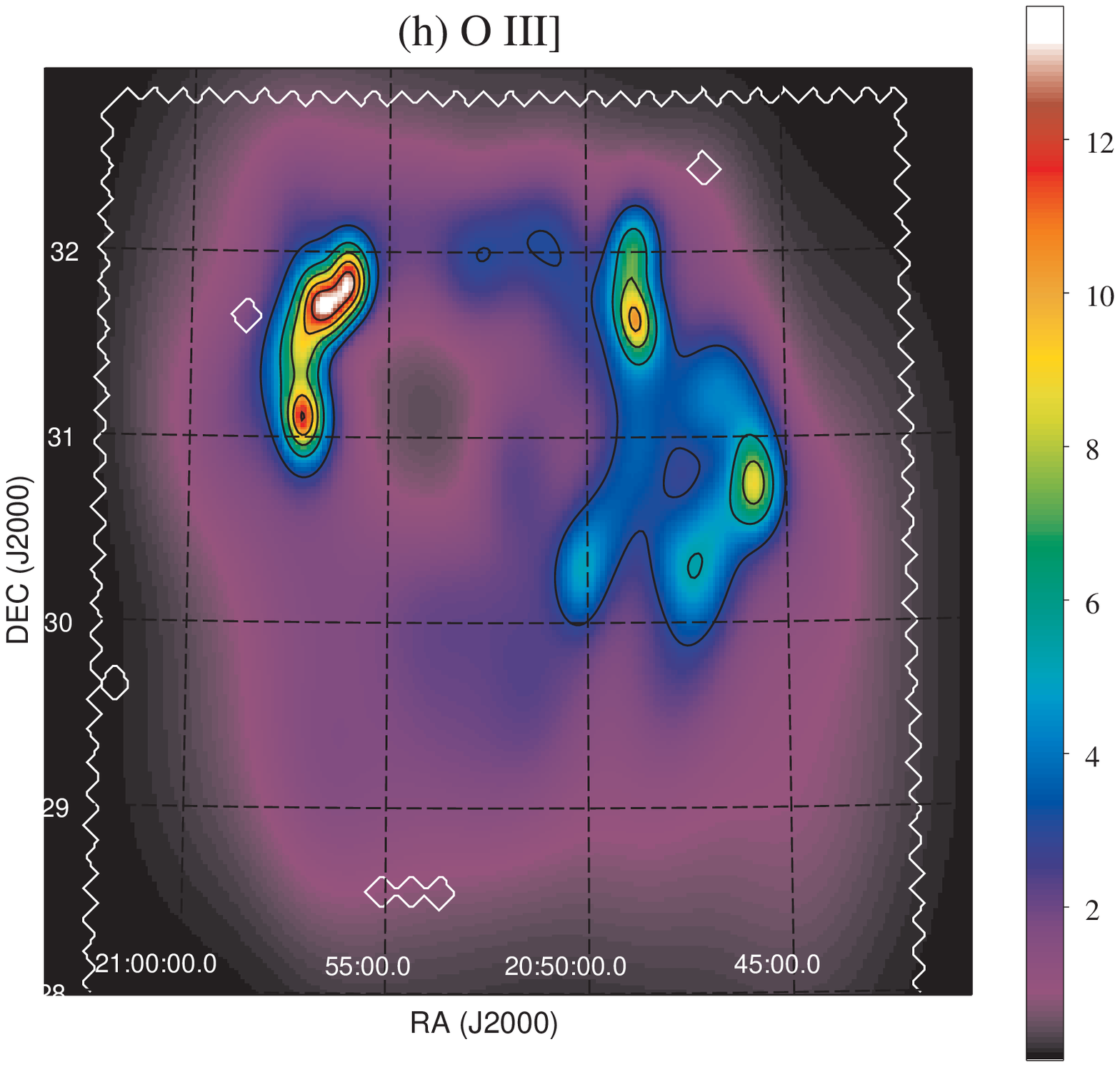}
\par\end{centering}
\textbf{Figure 1.} SPEAR/FIMS (a) \ion{C}{3} $\lambda$977, (b) \ion{O}{6} $\lambda\lambda$1032, 1038, (c) \ion{Si}{4} $\lambda\lambda$1394, 1403, (d) \ion{O}{4}{]} $\lambda$1404, (e) \ion{N}{4}{]} $\lambda$1486, (f) \ion{C}{4} $\lambda\lambda$1548, 1551, (g) \ion{He}{2} $\lambda$1640.5, and (h) \ion{O}{3}{]} $\lambda\lambda$1661, 1666 emission-line images of the Cygnus Loop (in units of 10$^{-6}$ erg s$^{-1}$ cm$^{-2}$ sr$^{-1}$). Contour levels are stepped by seven equal intervals and range (a) from 15.0 to 36.0, (b) from 25.0 to 70.0, (c) from 1.3 to 6.1, (d) from 1.5 to 9.5, (e) from 1.2 to 4.2, (f) from 3.5 to 20.0, (g) from 3.0 to 12.0, and (h) from 3.0 to 12.0. The solid lines indicate the outlines of the masked pixels.
\end{figure}

Figure 1 shows the resulting eight emission-line images. The solid lines denote the masked bright pixels. The images have been adaptively smoothed using a Gaussian function with a position-variable kernel radius, which is the same as the adaptive kernel method adopted in \citet{seon06}. The smoothing-kernel scales of $\sim$3$\arcmin$ to $\sim$30$\arcmin$ were used to obtain moderate signal-to-noise ratios (S/N). Seven equally-spaced contour levels were overlaid on each image, and all pixels inside the lowest-level contours have S/N $>$3. Figures 1(a) and (b) have lower spatial resolutions than the other images. This is because the \ion{C}{3} and \ion{O}{6} lines belong to the S-channel (with relatively lower sensitivity) and consequently, the images have been smoothed with larger smoothing-kernel scales. Moreover, the eastern region has much lower exposure time for the S-channel observation, which may make the \ion{C}{3} and \ion{O}{6} local peaks in the eastern region weaker and more diffuse. Figures 1(c) and (d) show a clear distinction between the deblended \ion{Si}{4} and \ion{O}{4}{]} images. In particular, the \ion{Si}{4} emission line is found to be strongest around a local peak on the eastern edge, but relatively weak around the \ion{O}{4}{]} peaks in the western region. Even though the \ion{N}{4}{]} line was weakest in the L-channel, we could obtain an \ion{N}{4}{]} emission-line image in Figure 1(e), which is compatible with the other emission-line images. The strongest \ion{N}{4}{]} peak appears to be around the \ion{Si}{4} peak on the eastern edge. In Figures 1(f)--(h), the \ion{C}{4}, \ion{He}{2}, and \ion{O}{3}{]} emission-line images have been reproduced using the newly processed data, and do not show large differences from the images by \citet{seon06}.

\begin{figure}[t]
\begin{centering}
\includegraphics[scale=0.22]{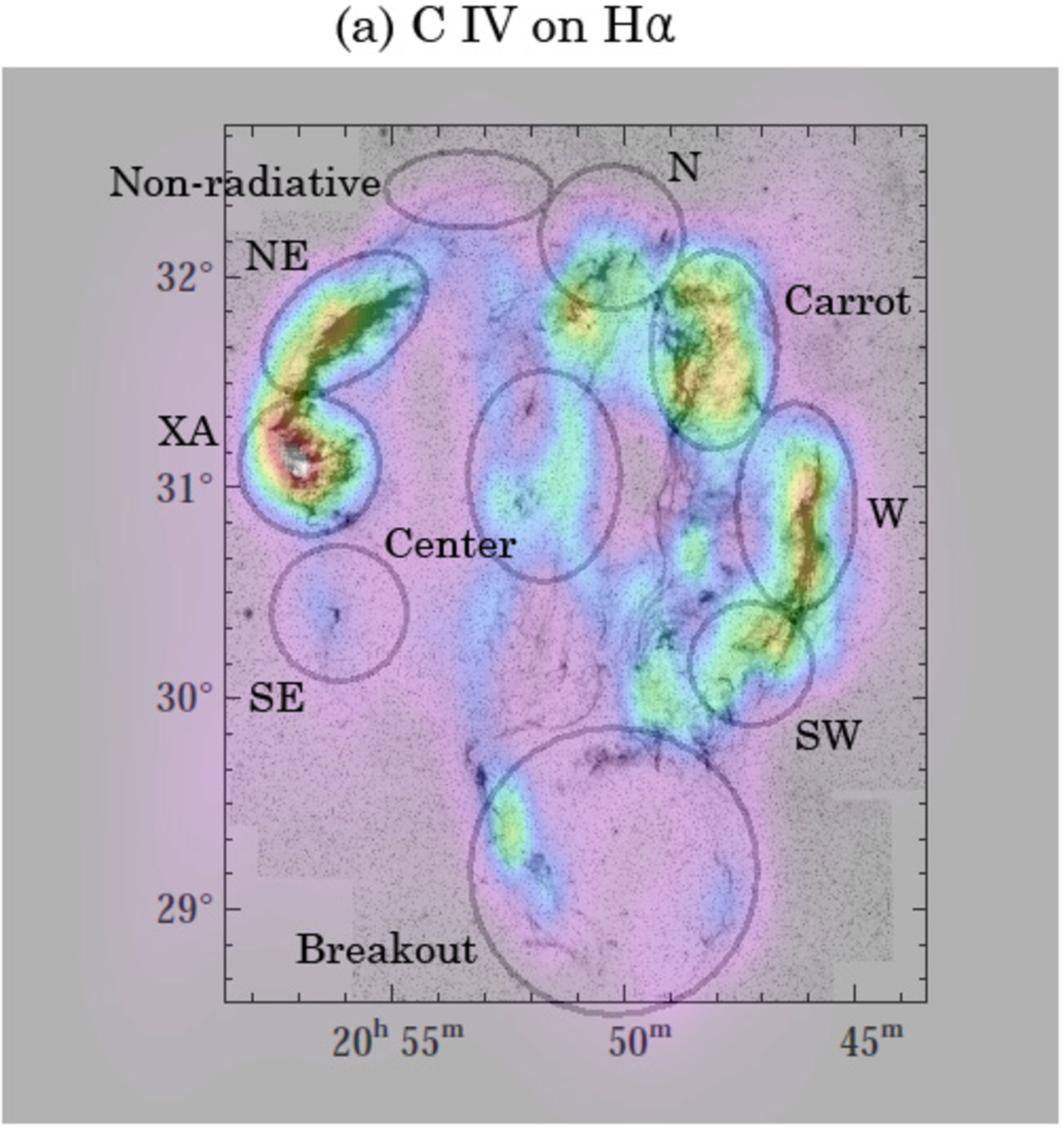}\quad\includegraphics[scale=0.22]{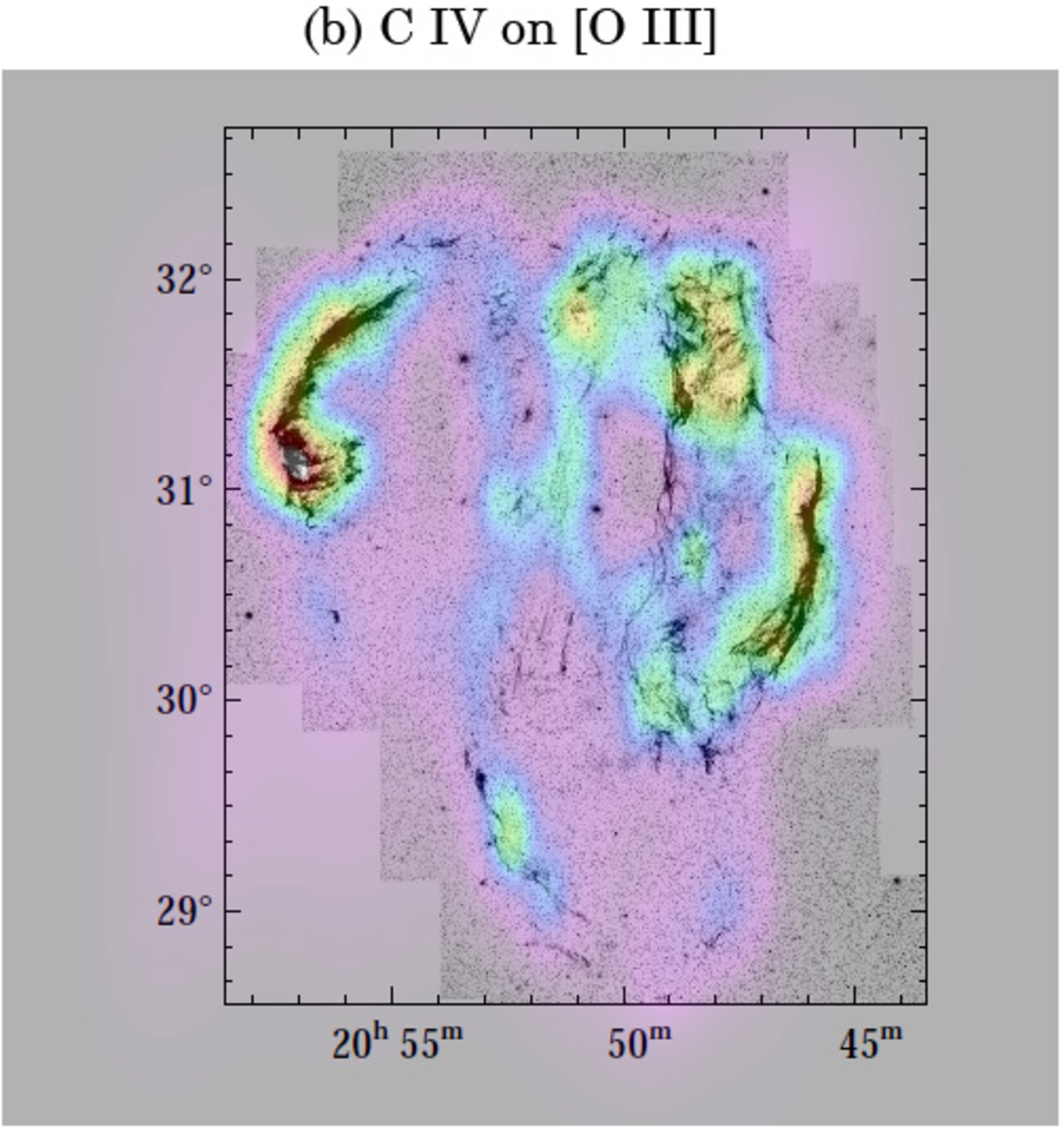}
\par\end{centering}
\vspace{0.1cm}
\begin{centering}
\includegraphics[scale=0.22]{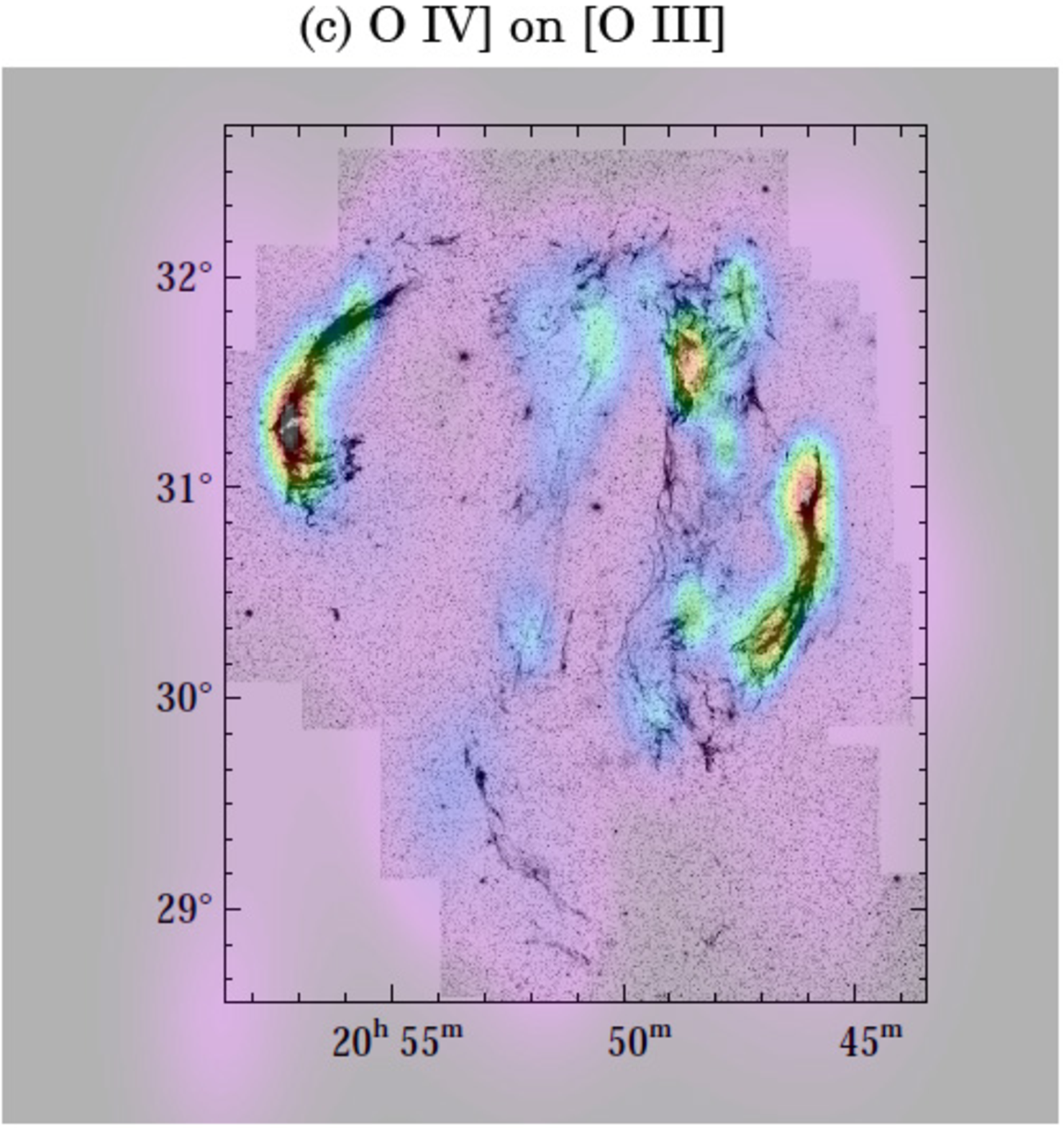}\quad\includegraphics[scale=0.22]{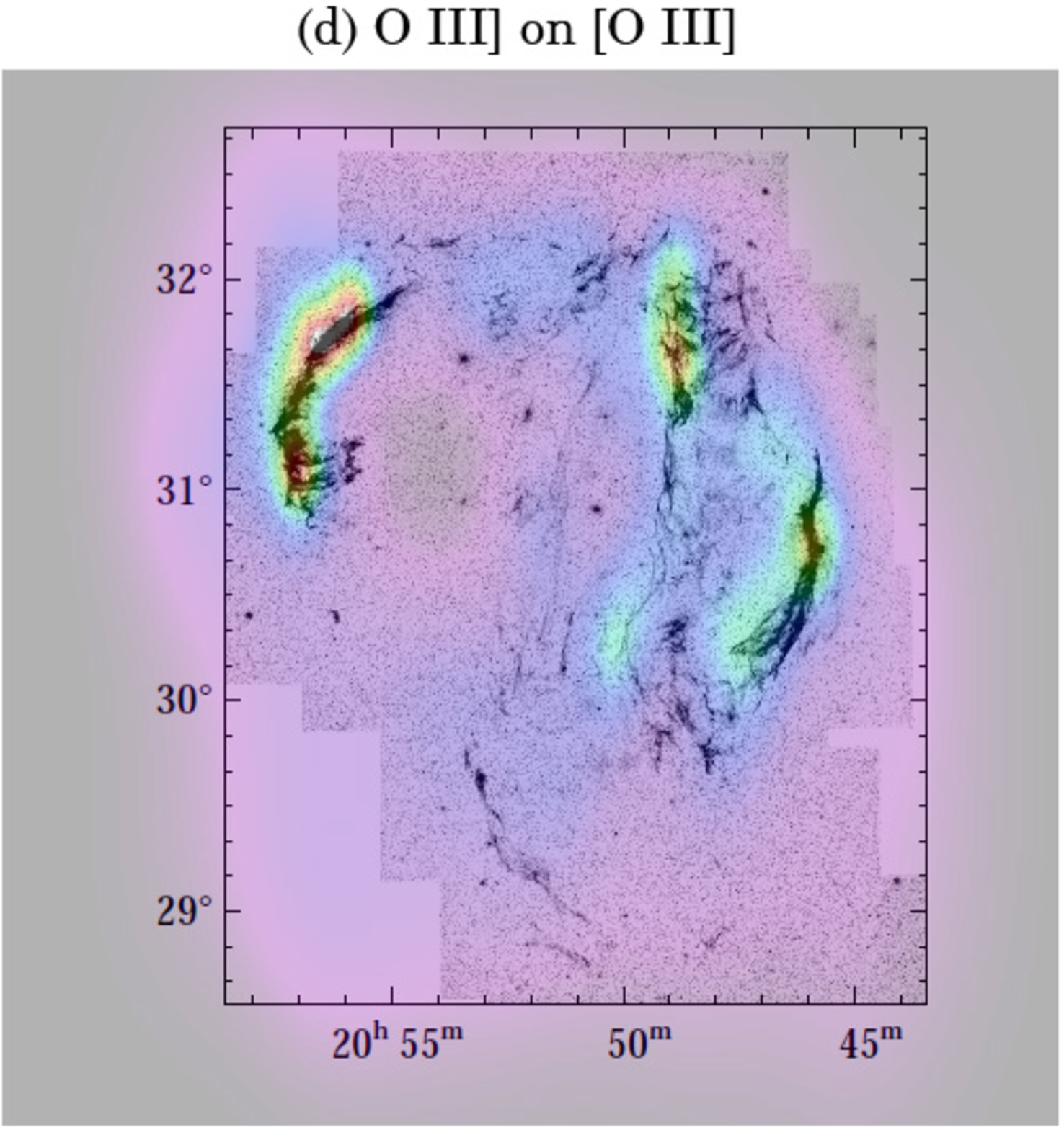}
\par\end{centering}
\vspace{0.1cm}
\begin{centering}
\includegraphics[scale=0.22]{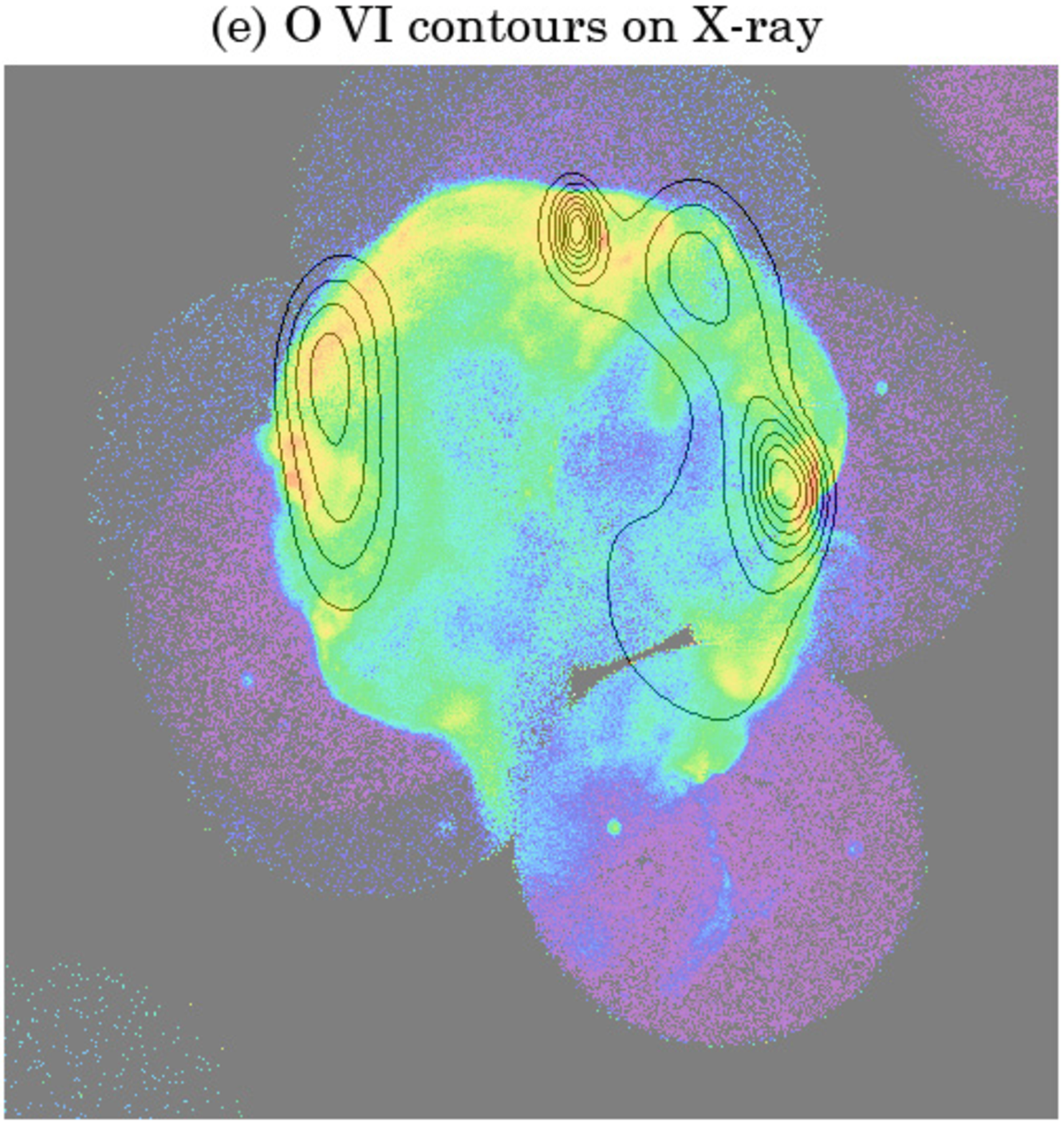}\quad\includegraphics[scale=0.22]{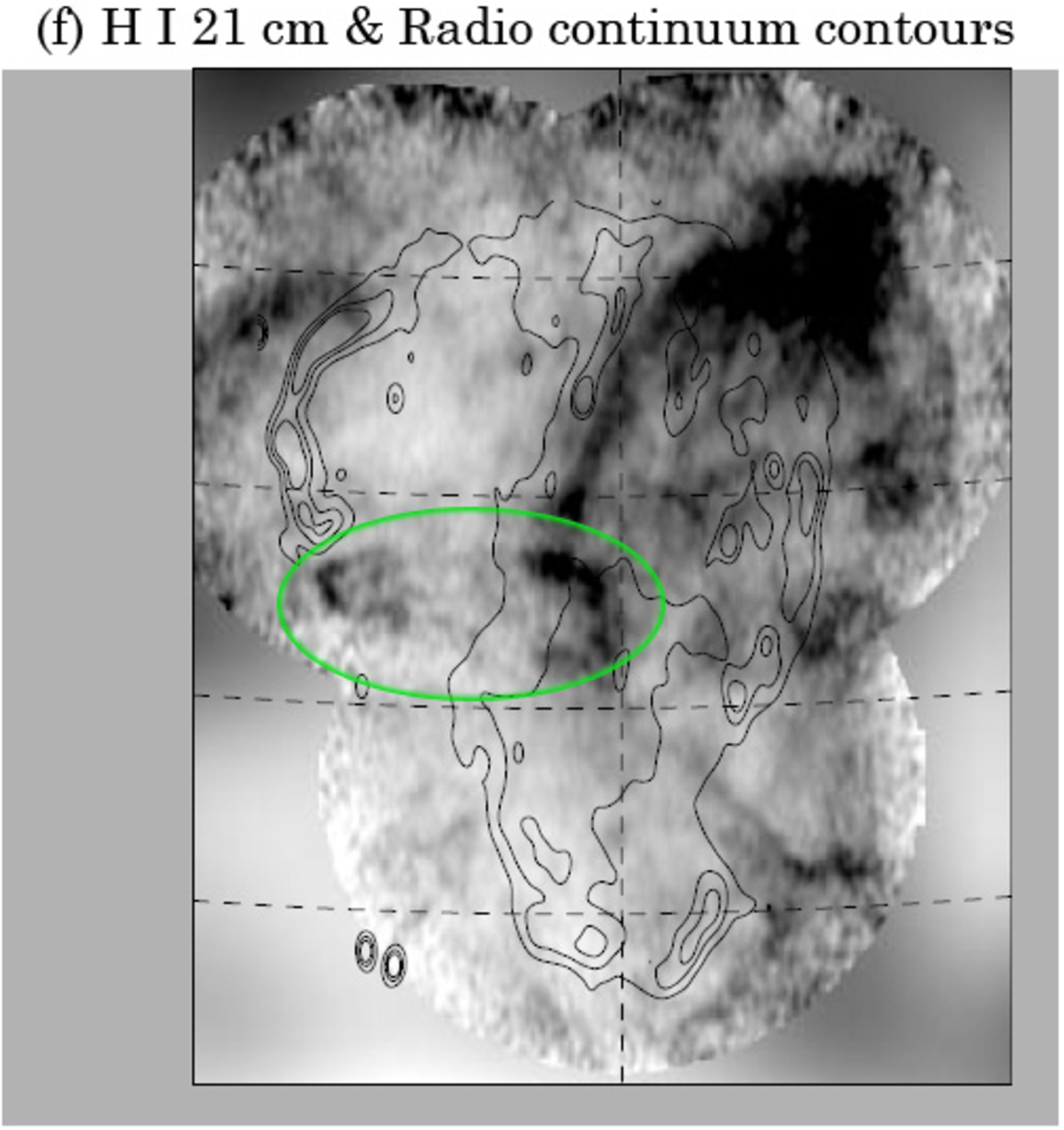}
\par\end{centering}
\textbf{Figure 2.} (a) \ion{C}{4} $\lambda\lambda$1548, 1551 color-scale image overlaid on H$\alpha$ grey-scale image. (b) \ion{C}{4}, (c) \ion{O}{4}{]} $\lambda$1404, and (d) \ion{O}{3}] $\lambda\lambda$1661, 1666 color-scale images overlaid on [\ion{O}{3}] $\lambda$5007 grey-scale images. The color-scale images are the same as in Figure 1, and the grey-scale images are from \citet{levenson98}. The ellipses marked in (a) denote subregions discussed in the text. (e) \ion{O}{6} $\lambda\lambda$1032, 1038 contours overlaid on {\it ROSAT} PSPC X-ray (0.1--2.4 keV) intensity image and (f) \ion{H}{1} 21 cm ($v_{\mathrm{LSR}}$ = -8.4 to 1.5 km s$^{-1}$) image with the 1420 MHz radio continuum contours \citep{leahy02}. The ellipse marked in (f) indicates the \ion{H}{1} ring feature identified by \citet{leahy05}.
\end{figure}

The SPEAR/FIMS FUV images are compared with the optical, X-ray, and radio images in Figure 2. In Figures 2(a)--(d), the \ion{C}{4}, \ion{O}{4}{]}, and \ion{O}{3}{]} images are overlaid onto the H$\alpha$ and [\ion{O}{3}] $\lambda$5007 grey-scale images adapted from \citet{levenson98}. The figures show strong correlations between the FUV images and optical images, although their spatial resolutions are fairly different. The \ion{C}{4} (the strongest-detected FUV emission line) closely matches both optical images, not only in prominent features but also in faint features, as can be seen in Figures 2(a) and (b). Ten subregions of the characteristic features, which will be discussed later, are denoted in Figure 2(a). In Figure 2(e), the \ion{O}{6} contours of Figure 1(b) are overlaid onto the {\it ROSAT} PSPC X-ray (0.1--2.4 keV) intensity image plotted using the data from the SkyView virtual observatory \citep{mcglynn98}. Although the \ion{O}{6} image has too low spatial resolution for detailed comparison, the \ion{O}{6} clearly appears to peak around the X-ray peaks on the eastern, northern, and western edges. Additionally, the \ion{H}{1} 21 cm image overlapped with the 1420 MHz radio continuum contours is shown in Figure 2(f), which is adapted from \citet{leahy02}. The \ion{H}{1} elliptical ring feature just inside the marked ellipse, is the identified \ion{H}{1} cloud located on the front side of the Cygnus Loop \citep{leahy05}. We note that all of the SPEAR/FIMS FUV images are faint in the direction of this \ion{H}{1} cloud, as shown in Figures 1 and 2(a)--(d).

\begin{figure}[t]
\begin{centering}
\includegraphics[scale=0.4]{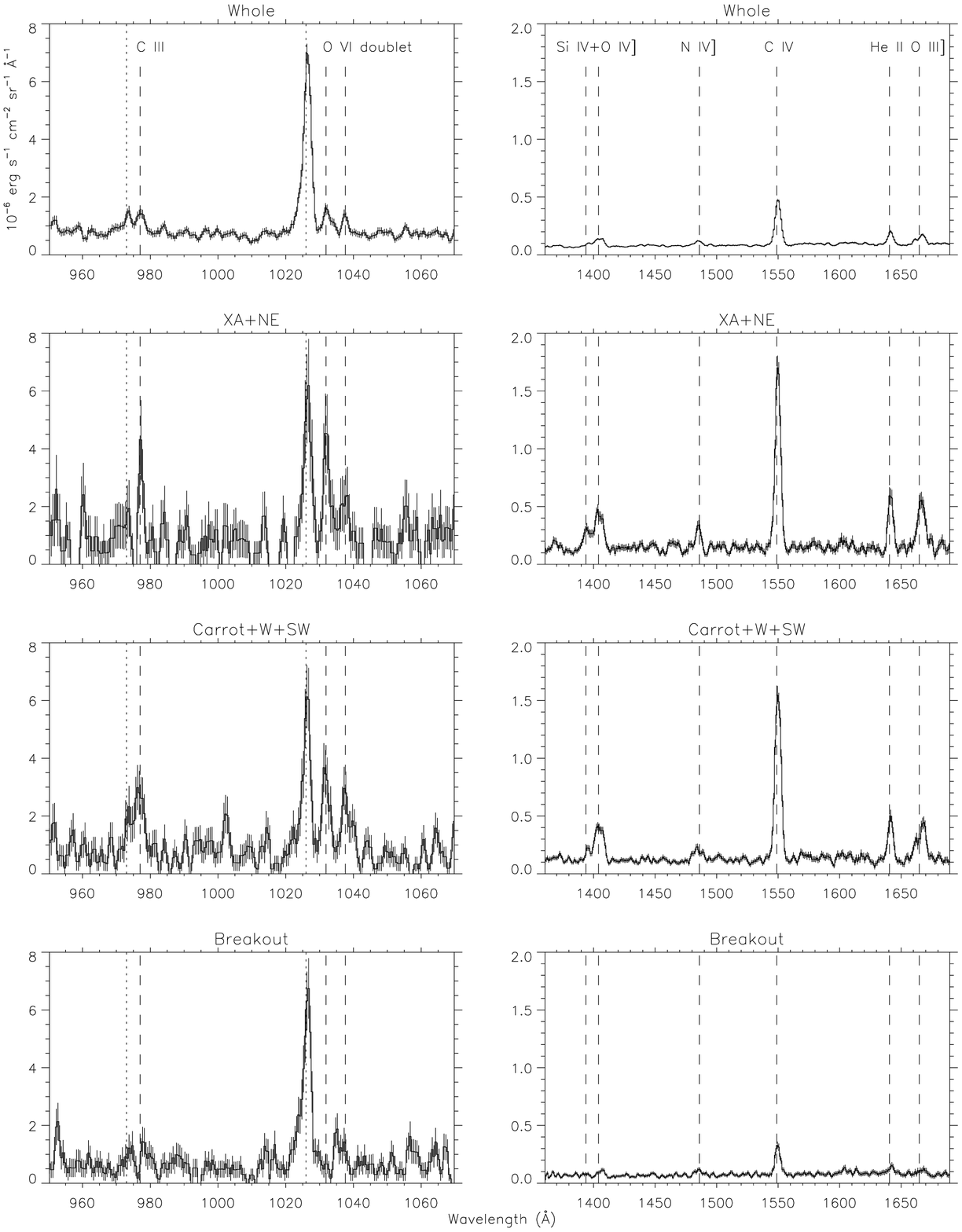}
\par\end{centering}
\textbf{Figure 3.} SPEAR/FIMS S-channel ($\it left$) and L-channel ($\it right$) spectra (with 1 $\sigma$ error bars) for the whole region and several subregions indicated in Figure 2(a). The spectra are binned with 0.5 (S-channel) and 1 (L-channel) \AA{} intervals, and smoothed with a boxcar average of 3 bins. Dashed lines denote the positions of the identified emission lines, and dotted lines indicate those of the hydrogen Lyman series ($\gamma$ 973 \AA{}, $\beta$ 1026 \AA{}) originating from the geocorona.
\end{figure}

In Figure 3, the SPEAR/FIMS FUV spectra for several subregions are plotted along with the spectra of the whole region. The detector background estimated in \citet{seon11} was subtracted from the spectra. The S-channel spectra ascertain the \ion{C}{3} $\lambda$977 and \ion{O}{6} $\lambda\lambda$1031.9, 1037.6 emission lines, together with the hydrogen Lyman series at 973 and 1026 \AA{} originating from the geocoronal airglow. The L-channel spectra show the \ion{Si}{4} $\lambda$1393.8, \ion{O}{4}{]} $\lambda$1404, \ion{N}{4}{]} $\lambda$1486, \ion{C}{4} $\lambda\lambda$1548, 1551 (unresolved), \ion{He}{2} $\lambda$1640.5, and \ion{O}{3}{]} $\lambda\lambda$1661, 1666 emission lines, all of which have already been reported in \citet{seon06}. However, the \ion{Si}{4} and \ion{O}{4}{]} lines, unresolved in \citet{seon06}, are clearly distinguishable in our spectra; particularly for subregions XA+NE and Carrot+W+SW. The spectra were fitted to calculate line intensities using the same method as for the emission-line maps. The line intensities were reddening-corrected by adopting the extinction curve of \citet{cardelli89} with $R_V$ = 3.1, and $E(\bv)$ = 0.08 \citep{miller74}. Although the extinction curve was provided only for wavelengths $>$1000 \AA{}, we extrapolated from the curve down to the wavelength of \ion{C}{3} $\lambda$977. In this short wavelength domain, other known extinction curves are also highly uncertain. Assuming a distance of 540 pc to the Cygnus Loop \citep{blair05}, the line luminosities were also calculated and are shown along with the values for the Vela SNR \citep{kim12} in Table 1. The X-ray luminosity \citep{ku84} is also given for comparison. The distance of 540 pc is slightly larger than that (440 pc) adopted in \citet{seon06}, and is a more recent, revised value \citep{blair05}. However, we note that there is some evidence that the distance to the Cygnus Loop may be larger than 540 pc \citep{salvesen09}. The SPEAR/FIMS \ion{C}{3}, \ion{O}{6}, and \ion{N}{4}{]} line luminosities of the Cygnus Loop have been newly obtained and the \ion{Si}{4} and \ion{O}{4}{]} line luminosities have been separately estimated in this study.

\begin{figure}[t]
\begin{centering}
\includegraphics[scale=0.22]{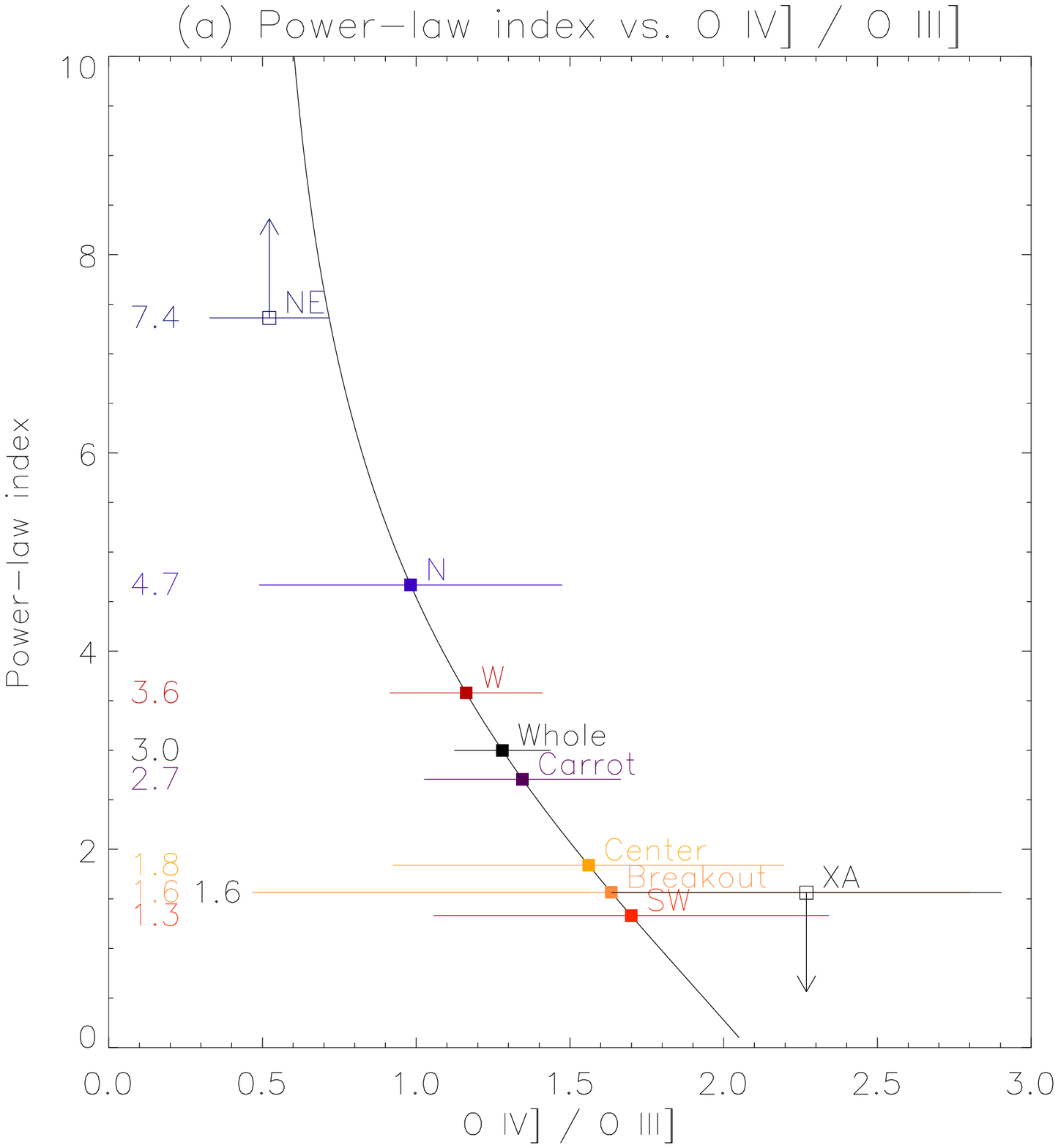}\quad\includegraphics[scale=0.22]{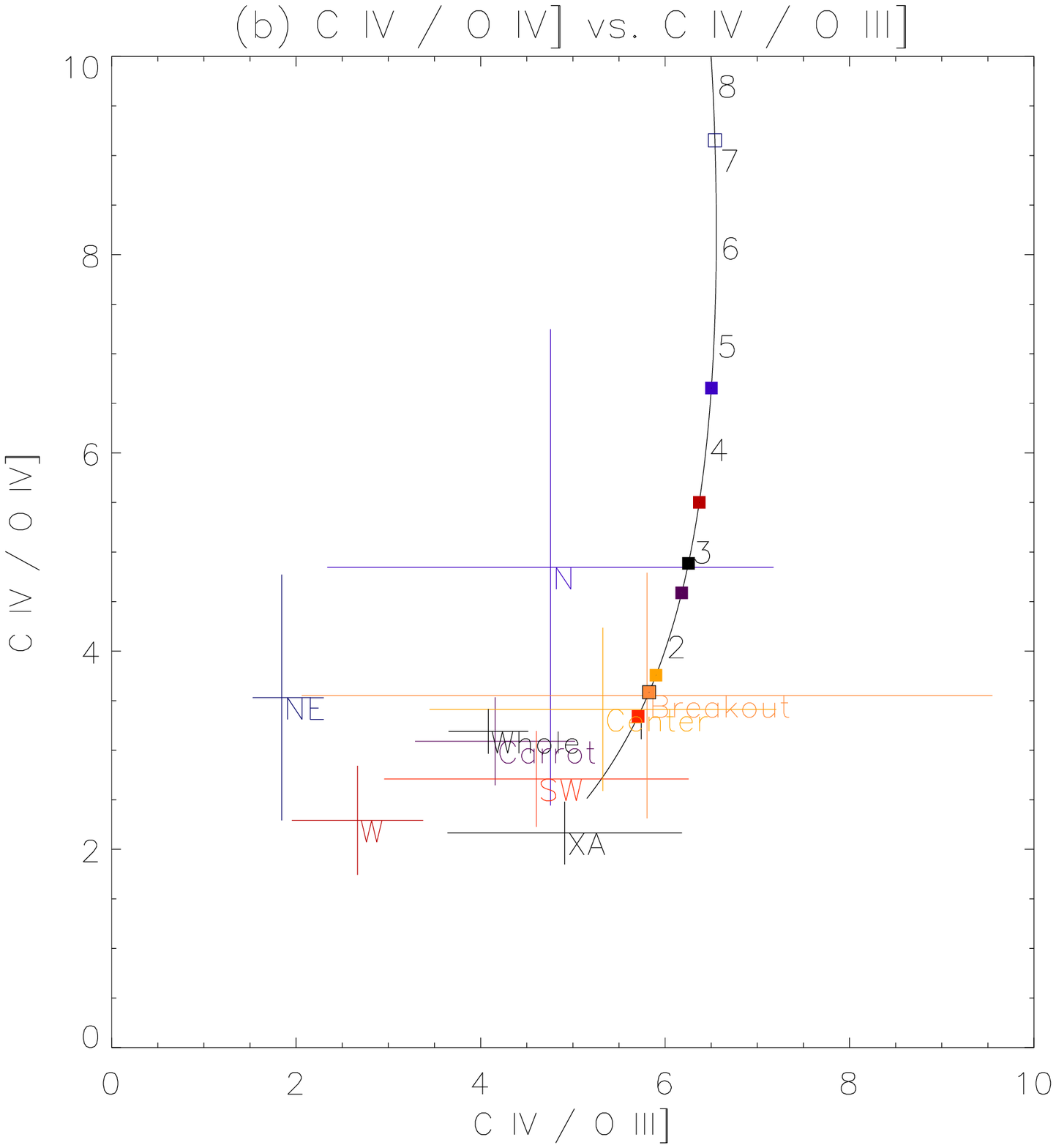}
\par\end{centering}
\vspace{0.3cm}
\begin{centering}
\includegraphics[scale=0.22]{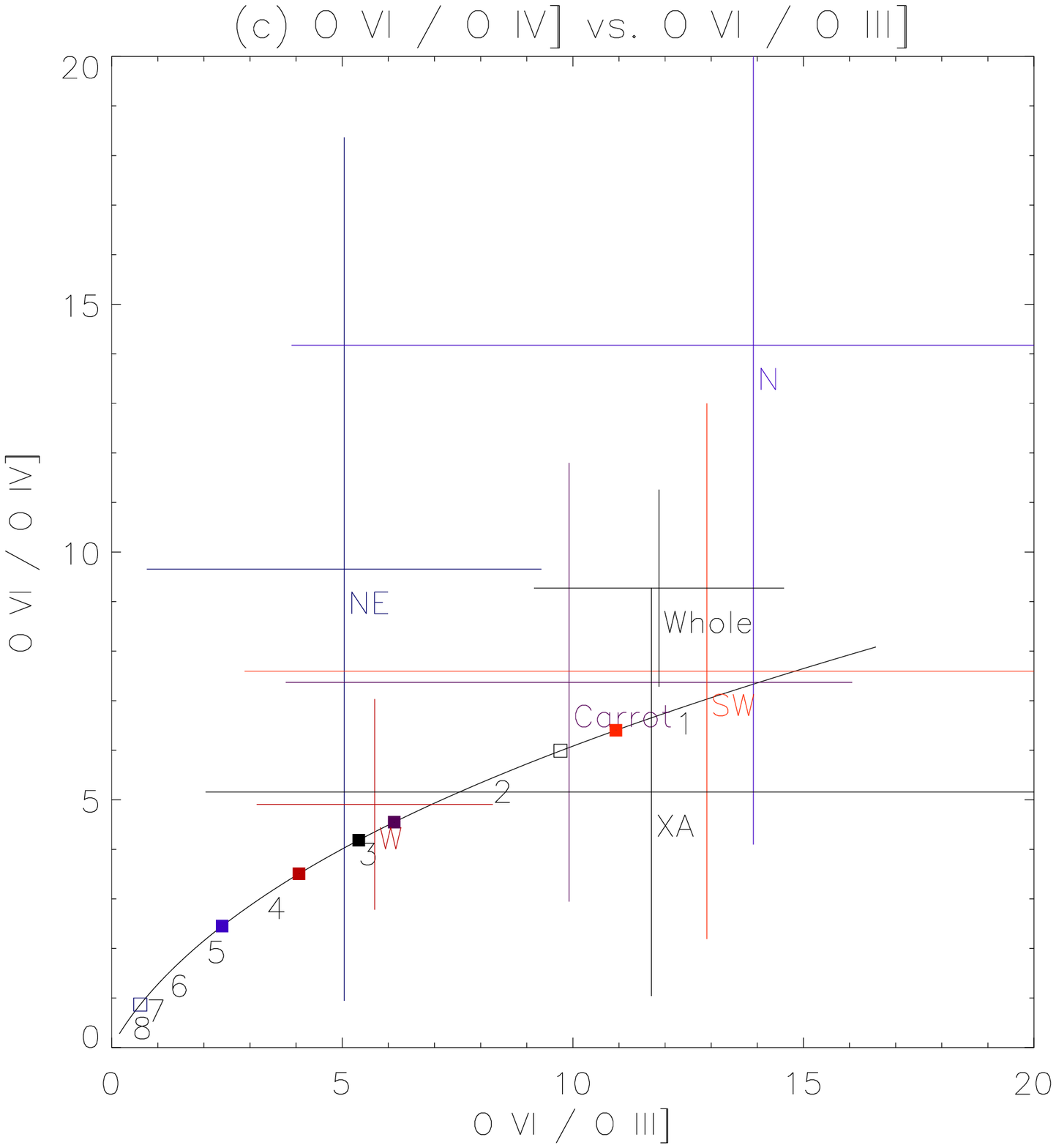}\quad\includegraphics[scale=0.22]{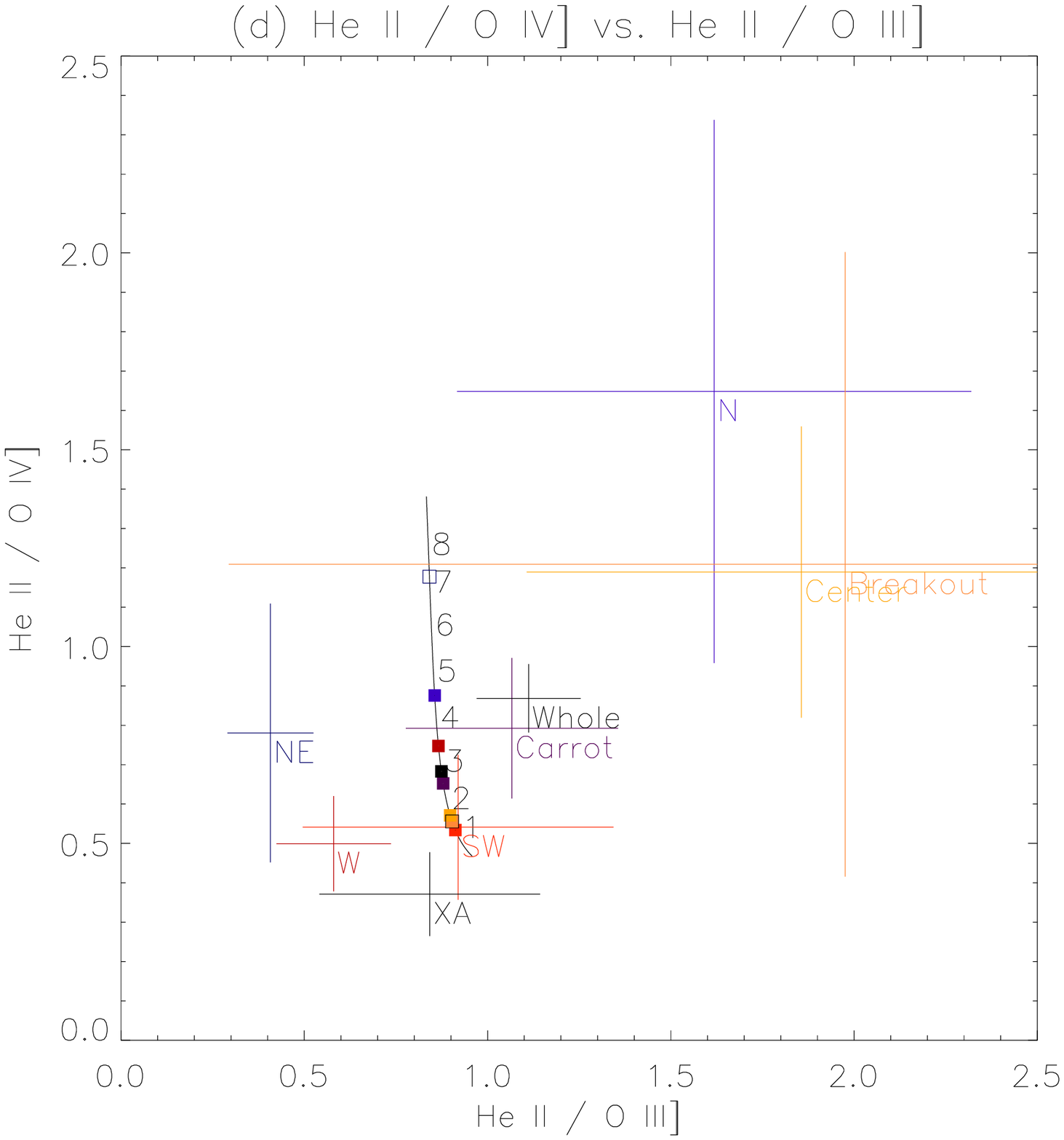}
\par\end{centering}
\textbf{Figure 4.} Line-ratio diagrams for a shock model and observed values. The curves indicate the line-ratio values expected from mixture of shocks with velocity populations of 0.1--10 power-law indices. The line ratios (with 1 $\sigma$ error bars) observed for the whole region and each subregion indicated in Figure 2(a) are overplotted. The line ratios less than 1 $\sigma$ error are not displayed. The values of the power-law indices to explain the observed \ion{O}{4}{]}/\ion{O}{3}{]} ratios in each subregion are denoted by numerical values on the left side in (a), and the corresponding positions (along with some integer values) are marked on the curves in (b)--(d).
\end{figure}

\begin{deluxetable*}{lccccccc}
\tabletypesize{\footnotesize}
\tablewidth{0pt}
\tablecaption{FUV line luminosities of the Cygnus Loop and comparison with the Vela SNR}
\tablehead{\colhead{Species} & \colhead{XA+NE} & \colhead{Carrot+W+SW} & \colhead{Breakout} & \colhead{Whole Cygnus Loop} & \colhead{Vela SNR\tablenotemark{a}}}
\startdata
\ion{C}{3} $\lambda$977                  & $<$1.64         & 3.44 $\pm$ 1.00 & $<$1.81         & 12.34 $\pm$ 2.55 & 21.84 $\pm$ 0.49 \\
\ion{O}{6} $\lambda\lambda$1032, 1038    & 2.95 $\pm$ 1.35 & 4.68 $\pm$ 1.35 & $<$1.54         & 19.79 $\pm$ 4.04 & 14.81 $\pm$ 0.45 \\
\ion{Si}{4} $\lambda\lambda$1394, 1403   & 0.27 $\pm$ 0.05 & $<$0.12         & $<$0.07         & 0.41 $\pm$ 0.14  & \nodata          \\
\ion{O}{4}{]} $\lambda$1404              & 0.45 $\pm$ 0.06 & 0.67 $\pm$ 0.06 & 0.10 $\pm$ 0.03 & 2.10 $\pm$ 0.14  & 6.13 $\pm$ 0.17  \\
\ion{N}{4}{]} $\lambda$1486              & 0.15 $\pm$ 0.03 & 0.17 $\pm$ 0.05 & 0.10 $\pm$ 0.03 & 0.79 $\pm$ 0.11  & 1.47 $\pm$ 0.07  \\
\ion{C}{4} $\lambda\lambda$1548, 1551    & 1.23 $\pm$ 0.07 & 1.90 $\pm$ 0.08 & 0.36 $\pm$ 0.04 & 6.73 $\pm$ 0.17  & 20.28 $\pm$ 0.12 \\
\ion{He}{2} $\lambda$1640.5              & 0.31 $\pm$ 0.05 & 0.48 $\pm$ 0.05 & $<$0.19         & 1.83 $\pm$ 0.14  & 2.75 $\pm$ 0.10  \\
\ion{O}{3}{]} $\lambda\lambda$1661, 1666 & 0.51 $\pm$ 0.06 & 0.58 $\pm$ 0.07 & $<$0.10         & 1.64 $\pm$ 0.17  & 3.56 $\pm$ 0.12  \\
X-ray                                    & \nodata         & \nodata         & \nodata    & 5.41\tablenotemark{b} & 3.0              \\
\enddata
\tablecomments{All line luminosities are in units of 10$^{35}$ ergs s$^{-1}$. The values for the Cygnus Loop are calculated adopting a distance of 540 pc \citep{blair05}, and reddening-corrected assuming $E(\bv)$ = 0.08 \citep{miller74}. Line luminosities obtained with $<$2 $\sigma$ significance are indicated by 1 $\sigma$ upper limits.}
\tablenotetext{a}{The values are from \citet{kim12} and the corresponding energy band for the X-ray is 0.1--2.5 keV.}
\tablenotetext{b}{The value is from \citet{ku84} and scaled to a distance of 540 pc. The corresponding energy band is 0.1--4.0 keV.}
\end{deluxetable*}

Since the SPEAR/FIMS data have limited spatial resolution, it would be impossible to estimate the velocity of an individual shock from the observed FUV line ratios. Instead, we attempted to compare the ratios of FUV lines summed over each subregion with the values expected from mixture of radiative shocks with a power-law velocity population \citep{vancura92}. The line ratios predicted for each shock velocity were obtained using the shock+precursor model of the MAPPINGS III code \citep{allen08}. We used the models with a preshock density of 1.0 cm$^{-3}$ and a preshock magnetic field of 0.5 $\mu$G, both of which do not significantly affect the results. We also used the atomic abundance set ``dopita2005'' (with $\sim$0.2--0.3 solar abundance for C, N, and O; Dopita et al. 2005) as defined in Table 1 of \citet{allen08} to reflect the low abundance in the rim of the Cygnus Loop \citep{katsuda08,leahy04,levenson05,uchida09}. Assuming a mixture of shocks with the velocity population P(V) $\propto$ V$^{-n}$, where shock velocities V are in the range of 100--250 km s$^{-1}$ with an interval of 25 km s$^{-1}$, we tried to determine the power-law indices ($n$), which match the observed line ratios. Here, the maximum shock velocity was selected to exclude non-radiative shocks. With a higher power-law index, the mixture of shocks contains a relatively large fraction of slower shocks. In Figure 4, we plot the line-ratio curves obtained by varying the power-law indices in the 0.1--10 range. The spectra extracted from the whole region and the subregions indicated in Figure 2(a), were used to estimate each line ratio and the resulting line ratios with 1 $\sigma$ error bars are overplotted. We present only the ratios for the \ion{O}{6}, \ion{C}{4}, \ion{He}{2}, \ion{O}{3}{]}, and \ion{O}{4}{]} lines. The \ion{C}{3} line was excluded because of high uncertainty regarding the reddening-correction, and the \ion{Si}{4} and \ion{N}{4}{]} lines were too weak to be reliable in most subregions. In Figure 4(a), the \ion{O}{4}{]} to \ion{O}{3}{]} ratio for subregion NE is well matched with the highest power-law index ($>$7.4 regarding 1 $\sigma$ error); indicating a relatively larger fraction of slow shocks in the subregion than in the other subregions. Subregions W and Carrot have the power-law indices of $\sim$3.6 and $\sim$2.7, respectively. The lowest index ($<$1.6 regarding 1 $\sigma$ error), meaning a large fraction of fast shocks, is found in subregion XA. The whole region has the intermediate value of $\sim$3.0, which seems to be a reasonable mixture of the subregions. However, the values of the power-law indices obtained from the \ion{O}{4}{]} to \ion{O}{3}{]} ratio do not coincide well with the results obtained with the \ion{C}{4}, \ion{O}{6}, or \ion{He}{2} lines, as can be seen in Figures 4(b)--(d). We will discuss the implications of these results in the next section.

\section{DISCUSSION}

Using the S-channel data of SPEAR/FIMS, we newly created the \ion{C}{3} $\lambda$977 and \ion{O}{6} $\lambda\lambda$1032, 1038 emission-line images of the Cygnus Loop in Figures 1(a) and (b). These images can be compared with the two previous results: the {\it Voyager 2} \ion{C}{3} and \ion{O}{6} images \citep{blair91b}, and the HIRES \ion{O}{6} contour map \citep{rasmussen92}. The SPEAR/FIMS \ion{C}{3} image has the brightest region along the northwestern side, and less bright regions on the eastern and northeastern sides. The brightest region coincides well with the brightest one in the {\it Voyager 2} \ion{C}{3} image. On the other hand, there is no feature corresponding to the eastern bright patch of the {\it Voyager 2} \ion{C}{3} image. Instead, a very diffuse feature is found in our \ion{C}{3} image in the vicinity of the bright patch. This difference could be due to a large smoothing scale in the eastern region of our image, as mentioned in Section 3, and/or the slit sampling effects of the {\it Voyager 2} data. The SPEAR/FIMS \ion{O}{6} image shows bright regions along the eastern side and the rims from the north to the southwest. Except for a faint region on the northeastern side and a somewhat diffuse, bright region on the eastern side, overall bright features agree well with those of the {\it Voyager 2} \ion{O}{6} image. The morphology of the \ion{O}{6} image can also be compared with the HIRES \ion{O}{6} contour map in more detail. A local peak on the western edge, and the eastern and northwestern enhanced regions, coincide with the HIRES \ion{O}{6} contour peaks. However, another local peak seen on the northern edge has no HIRES counterpart, although it seems to be near the northern hot spot of the {\it Voyager 2} image noted in \citet{blair91b}. With a slightly better resolution, the SPEAR/FIMS \ion{C}{3} and \ion{O}{6} images are not affected by the slit sampling effects of the {\it Voyager 2} data. Our \ion{O}{6} image shows better correlation with other wavelength images than the HIRES \ion{O}{6} contour map does. This will be further discussed later.

We note that our new \ion{C}{4}, \ion{He}{2}, and \ion{O}{3}{]} images in Figures 1(f)--(h) are not exactly the same as those in \citet{seon06}. These differences are mainly due to differences in the data set and the line-fitting methods (e.g., a linear versus a constant continuum, and the intensity ratios of doublet lines). Most of our results for subregions indicated in Figure 2(a) are similar to those of \citet{seon06}, and thus we discuss only the distinctions and new results. \citet{seon06} reported the detection of \ion{C}{4} and \ion{He}{2} originating from the northeastern, non-radiative region. However, selecting a region with a slightly higher declination than the NE non-radiative region in \citet{seon06}, as shown in Figure 2(a), results in the \ion{C}{4} and \ion{He}{2} detections with a significance $<$2 $\sigma$: just 1 $\sigma$ upper limits of $<$0.42 and $<$1.31 $\times$ 10$^{-6}$ erg s$^{-1}$ cm$^{-2}$ sr$^{-1}$, respectively. According to Figure 7 in \citet{levenson98}, there are incomplete radiative shocks located just below the southern border of subregion Non-radiative marked in Figure 2(a). Considering the spatial resolution of the SPEAR/FIMS data, most of the \ion{C}{4} and \ion{He}{2} detections reported in \citet{seon06} may not originate from non-radiative shocks but from incomplete radiative shocks. Even the \ion{O}{6} line, which is known to be emitted more strongly from non-radiative shocks, has not been detected in subregion Non-radiative. \citet{raymond83} predicted the \ion{O}{6} intensity of 1.2--12 $\times$ 10$^{-6}$ erg s$^{-1}$ cm$^{-2}$ sr$^{-1}$ from a sharp northeastern non-radiative shock. If this intrinsic brightness is obscured by interstellar dust, then the intensity would become lower and it is likely impossible to detect by the SPEAR/FIMS instrument with the limited spatial resolution. Subregions XA and NE are the brightest regions for all of the SPEAR/FIMS FUV lines, except for the S-channel lines (\ion{C}{3} and \ion{O}{6}), which have worse statistics there. However, their peak positions show apparent disagreement, as evident in Figure 1. The \ion{C}{4}, \ion{O}{4}{]}, and \ion{N}{4}{]} lines are brightest in subregion XA, whereas the \ion{He}{2} and \ion{O}{3}{]} lines are found to be stronger in subregion NE. The \ion{Si}{4} line separated from the \ion{O}{4}{]} line peaks around the middle point between the two subregions. The newly observed \ion{Si}{4} and \ion{N}{4}{]} lines have been identified in subregions Carrot, W, and Breakout, as well.

For five 40$\arcmin$ diameter regions, \citet{danforth00} presented FUV images with the spatial resolution comparable to those of the optical H$\alpha$ and [\ion{O}{3}] $\lambda$5007 images. Their FUV images show very sharp features coinciding well with bright optical filaments, although they have wide bandpass including the \ion{N}{4}{]}, \ion{C}{4}, \ion{He}{2}, and \ion{O}{3}{]} lines. Therefore, despite the coarse spatial resolutions of the FUV images, strong correlations between the FUV images and optical images in Figures 2(a)--(d), imply that most of FUV emissions seen in the figure likely originate from the shocks consisting of bright H$\alpha$ and [\ion{O}{3}] filaments. Assuming this, we will compare the FUV images with the optical images. Since the \ion{C}{4} resonance line is formed at roughly the same temperature as the [\ion{O}{3}] $\lambda$5007 forbidden line \citep{raymond97}, Figure 2(b) could directly provide clues to the regions where strong resonance-scattering occurs. A vertical filamentary structure running from $\delta \sim 30\arcdeg45\arcmin$ to $\delta \sim 31\arcdeg15\arcmin$ at $\alpha \sim 20^{\mathrm h}49^{\mathrm m}$ in the [\ion{O}{3}] image seems to have no \ion{C}{4} counterpart. Therefore, this region may be regarded as a candidate region of strong-resonance-scattering. However, the vertical feature is also absent in the other FUV line images in Figure 1, except for a very faint feature in the \ion{O}{3}{]} image. Therefore, the absence of the vertical feature in the \ion{C}{4} image may not be due to the effect of resonance scattering, but instead be due to dust extinction by a foreground dust cloud. The optical, vertical filamentary feature would not be extinguished even by the foreground dust cloud because of much weaker dust extinction in optical wavelengths. This might be another case of the \ion{H}{1} cloud displayed in Figure 2(f), which has been considered to be on the front side of the Cygnus Loop, as noted in Section 3. Since the \ion{O}{4}{]}, \ion{O}{3}{]}, and [\ion{O}{3}] lines are all free from resonance scattering, and their relative strengths are not affected by elemental abundances, the comparison between Figures 2(c) and (d) can be used to verify spatial variations of shock-velocity populations. The obvious variations of the \ion{O}{4}{]} and \ion{O}{3}{]} intensities are found on two filamentary [\ion{O}{3}] features: one running along subregions XA and NE, and the other running along subregions W and SW. On the former feature, the \ion{O}{4}{]} intensity is highest around the northern part of subregion XA, whereas the \ion{O}{3}{]} intensity becomes higher toward both ends of the filamentary feature. On the latter feature, the \ion{O}{4}{]} intensity becomes higher toward both ends of the filamentary feature, but the \ion{O}{3}{]} intensity is highest near the center of subregion W. In the shock models of Figure 4(a), the line ratio \ion{O}{4}{]}/\ion{O}{3}{]} increases with the relative fraction of fast shocks. It should be noted that the positions with higher \ion{O}{4}{]}/\ion{O}{3}{]} on each filamentary feature are closer to the locations of bright X-ray peaks seen in Figure 2(e). This is compatible with the above argument about shock-velocity populations.

As mentioned above, Figure 2(e) shows close coincidence between the SPEAR/FIMS \ion{O}{6} and the X-ray peaks; particularly the northern and the western peaks. The bright region on the eastern side of the \ion{O}{6} image is well correlated with the overlapped X-ray peaks there. An extended, fainter region on the southwestern side also overlaps well with the V-shaped X-ray feature \citep{aschenbach99,leahy04}. The similarity between the \ion{O}{6} and X-ray images indicates that the observed \ion{O}{6} emissions are mostly originating from the X-ray emitting gas. The only enhanced \ion{O}{6} region without bright X-ray counterparts is located on the northwestern side, which overlaps with the northern part of the optical features in subregion Carrot. To estimate the amount of the contribution by X-ray emitting gas, we compared the observed \ion{O}{6} intensity with the value expected from the observed X-ray emissions. \citet{decourchelle97} calculated the \ion{O}{6} intensities predicted at the northern X-ray peak (their Field A) using a few models. Their EI + NEI model predicts a similar intensity (59 $\times$ 10$^{-6}$ erg s$^{-1}$ cm$^{-2}$ sr$^{-1}$) as the average intensity of $\sim$50 $\times$ 10$^{-6}$ erg s$^{-1}$ cm$^{-2}$ sr$^{-1}$ observed at the northern \ion{O}{6} peak in Figure 1(b). However, their two-EI model predicts only 8\% of the observed \ion{O}{6} intensity. These indicate that most of the \ion{O}{6} emissions near the northern peak can be interpreted to originate from the X-ray emitting gas of which the softer X-ray component is in the condition of equilibrium ionization, and the harder one is in the non-equilibrium ionization condition (EI + NEI model of \citet{decourchelle97}).

The FUV line luminosities in Table 1 are now compared with the previous results. For the \ion{C}{3} and \ion{O}{6} lines, we compare the observed intensities, which were not reddening-corrected, because of high uncertainty in the reddening-correction. \citet{blair91b} reported the average observed intensities of 4.0 $\times$ 10$^{-6}$ erg s$^{-1}$ cm$^{-2}$ sr$^{-1}$ for \ion{C}{3} and 8.8 $\times$ 10$^{-6}$ erg s$^{-1}$ cm$^{-2}$ sr$^{-1}$ for \ion{O}{6}. On the other hand, \citet{rasmussen92} obtained the \ion{O}{6} intensity of 3.0($\pm$1.2) $\times$ 10$^{-6}$ erg s$^{-1}$ cm$^{-2}$ sr$^{-1}$. Fitting the spectrum for the whole region in Figure 3 results in the observed intensities of 2.22($\pm$0.46) $\times$ 10$^{-6}$ erg s$^{-1}$ cm$^{-2}$ sr$^{-1}$ for \ion{C}{3} and 4.22($\pm$0.86) $\times$ 10$^{-6}$ erg s$^{-1}$ cm$^{-2}$ sr$^{-1}$ for \ion{O}{6}. These values are roughly half the {\it Voyager 2} values \citep{blair91b}. However, \citet{quemerais13} have recently recalibrated the {\it Voyager 2} intensities and found the old values to be too high by a factor of 2, meaning that our values are similar to the corrected {\it Voyager 2} values. The HIRES \ion{O}{6} value \citep{rasmussen92} also agrees well with our \ion{O}{6} intensity within 1 $\sigma$ error. We note that the SPEAR/FIMS systematic errors for the sensitivity calibration were estimated to be $\sim$25\% \citep{edelstein06b}. If the same distance as in \citet{seon06} is adopted, the present \ion{C}{4} luminosity for the whole region is almost exactly the same as that of \citet{seon06}. The \ion{He}{2} and \ion{O}{3}{]} luminosities, however, are $\sim$70--80\% higher than those of \citet{seon06}. The difference in the line-fitting methods, as mentioned above, likely caused the larger gaps for the weaker emission lines. \citet{seon06} assumed a fixed ratio of 1:5.5 for the blended \ion{Si}{4}+\ion{O}{4}{]} lines to calculate the \ion{O}{4}{]} to \ion{O}{3}{]} ratio. This caused a somewhat higher line ratio in subregion XA than those in the other parts of the Cygnus Loop. \citet{seon06} discussed that the \ion{O}{4}{]} to \ion{Si}{4} ratio in subregion XA would in fact, be lower than 5.5, based on the result of \citet{danforth01}. The decomposed line luminosities of \ion{Si}{4} and \ion{O}{4}{]} in Table 1 verify their argument. The \ion{O}{4}{]} to \ion{Si}{4} ratio for the whole region is $\sim$5.1, whereas the ratio for subregion XA+NE is $\sim$1.7. Fitting the spectrum extracted from only subregion XA, results in the ratio of $\sim$2.6. The relative increase of the \ion{Si}{4} line in subregion XA+NE can be recognized in the spectra of Figure 3.

In estimating shock velocities from comparisons between the observed FUV emission lines and radiative shock models, we should consider the effects of resonance scattering, X-ray emitting gas, and non-radiative shocks \citep{raymond97}. The \ion{O}{4}{]} to \ion{O}{3}{]} ratio is not affected by the above effects. Therefore, the fact that the values of the power-law indices obtained from the \ion{O}{4}{]} to \ion{O}{3}{]} ratio in Figure 4(a) are not derived well from the other line ratios, implies that the above effects are indeed important in the line ratios containing the \ion{C}{4}, \ion{O}{6}, and \ion{He}{2} lines. We, therefore, can infer the importance of the above effects in Figures 4(b)--(d). In Figure 4(b), the observed ratios \ion{C}{4}/\ion{O}{4}{]} and \ion{C}{4}/\ion{O}{3}{]} are found to be much lower than the values expected from the ratio \ion{O}{4}{]}/\ion{O}{3}{]} for most subregions. This indicates that the \ion{C}{4} intensities were significantly diminished by the resonance scattering, which is ubiquitous across the Cygnus Loop. The resonance scattering in subregions NE and W seems to be particularly strong (reduced by a factor of $>$3 and $>$2, respectively). Within our wide subregions, overlapping projections of the less edge-on shocks may cause a significant contribution from diffuse emission, in which the resonance scattering effect is less important. The contribution would depend on the nature of the ripples in the shock front. However, more numerous optical filaments in subregions NE and W than the other subregions, indicate the significance of the resonance scattering in subregions NE and W. A kinematic study of the whole Cygnus Loop region showed that the radial velocities around subregions NE and W are similar to the systemic velocity of the Cygnus Loop \citep{shull91}. Even the ratios for the whole region are reduced to $\sim$65\% of the expected values. Since the resonant scattering does not destroy photons, the decrease of the line ratios obtained from the whole region implies that the Cygnus Loop would neither be perfectly spherical nor be homogeneous. Another possibility would be the depletion onto dust grains. A global IR study of the Cygnus Loop showed that there is plenty of dust correlated well with the X-ray and optical emission of the remnant \citep{arendt92}. Also, some X-ray studies suggested that a possible explanation of the low abundance in the overall rim of the Cygnus Loop could be the dust depletion \citep{katsuda08,uchida09}. The difference in the degree of dust depletion for carbon and oxygen may affect the ratios \ion{C}{4}/\ion{O}{4}{]} and \ion{C}{4}/\ion{O}{3}{]}. As \citet{raymond03} presented clear evidence for resonance scattering of \ion{O}{6} in a northeastern non-radiative shock, resonance scattering of \ion{O}{6} can also be widespread throughout the Cygnus Loop. However, Figure 4(c) does not show any decrease in the line ratio for \ion{O}{6}. Instead, the line ratio seems to increase for some subregions, although their errors are somewhat large. The ratios for the whole region are higher by a factor of $>$2 than the values matching with the ratio \ion{O}{4}{]}/\ion{O}{3}{]}. These enhancements can be explained by the above conclusion that large portions of the observed \ion{O}{6} emissions would be due to the X-ray emitting gas. Moreover, since the X-ray emitting gas fills more diffuse and larger volumes inside the radiative shocks \citep{hester94}, the \ion{O}{6} emissions from the X-ray gas are less affected by resonance scattering. Figure 4(d) for the \ion{He}{2} line shows no clear tendency. Only in subregion N, ratios are found to be slightly higher than those expected. This may be due to a relatively large number of non-radiative shocks within subregion N because some portions of the \ion{He}{2} intensity can originate from non-radiative shocks \citep{hester94,long92,raymond83}.

In Table 1, the luminosities of the X-ray and the highest ionization FUV line (\ion{O}{6}) for the Cygnus Loop are $\sim$80\% and $\sim$30\% higher than those for the Vela SNR, respectively. On the other hand, the luminosities of the other FUV lines for the Cygnus Loop are all $\sim$30--70\% lower than those for the Vela SNR. We note that the ratios of the X-ray and \ion{O}{6} luminosities to those of the lower ionization lines are higher in the Cygnus Loop than in the Vela SNR. Although the distances to the Cygnus Loop was recently found to be larger than the value adopted in the present paper \citep{salvesen09}, the luminosity ratios are independent of the distance. Therefore, our result suggests that the Cygnus Loop must possess a much larger fraction of shocks in the early evolutionary stages than the Vela SNR. We conclude that the Cygnus Loop is globally younger than the Vela SNR from the point of view of shock evolution populations. In fact, the Vela SNR exploded near the border of the $\gamma^{2}$ Velorum stellar wind bubble and large portions of the remnant have already passed through the dense wall of the bubble \citep{kim12,sushch11}. In contrast, the Cygnus Loop exploded near the center of a cavity formed by its progenitor star and it has recently encountered the cavity walls \citep{hester94,levenson97}.

\section{CONCLUSIONS}

Utilizing the short wavelength channel data of SPEAR/FIMS, we have obtained new spectral images of the Cygnus Loop at \ion{C}{3} $\lambda$977 and \ion{O}{6} $\lambda\lambda$1032, 1038. These images supplement the previously reported images with the slit sampling effects or a little worse resolution. The observed intensities agree with the HIRES value within 1 $\sigma$ error. They are also consistent with the recently recalibrated {\it Voyager 2} values. In addition, the \ion{Si}{4}+\ion{O}{4}{]} lines unresolved in \citet{seon06} have been successfully decomposed. A new image for the weak \ion{N}{4}{]} $\lambda$1486 line was also obtained by binning and fitting the spectra carefully with the newly processed data. The deblended \ion{Si}{4} and \ion{O}{4}{]} line maps confirm a region with a relatively low \ion{O}{4}{]} to \ion{Si}{4} ratio, as predicted in \citet{seon06}.

We have compared the SPEAR/FIMS FUV images with the X-ray, optical, and \ion{H}{1} 21 cm images. The very close coincidence between the \ion{O}{6} and X-ray enhanced regions suggests that the observed \ion{O}{6} emissions are most likely related to X-ray emitting gas. For the region around the northern \ion{O}{6} peak, comparison of the observed \ion{O}{6} intensity with the value predicted from the observed X-ray emissions indicates that most of the \ion{O}{6} emissions observed there can be produced from the X-ray emitting gas. Comparisons with the optical H$\alpha$ and [\ion{O}{3}] $\lambda$5007 images have revealed clear variations of shock-velocity populations on the two filamentary features. We also argue that the FUV lines may be strongly extinguished in the direction of the previously identified \ion{H}{1} cloud, which would be located on the front side of the Cygnus Loop.

We have calculated the FUV line ratios for several subregions of the Cygnus Loop, and compared them with a radiative shock model. From this, we have diagnosed subregions containing relatively large fractions of slow or fast radiative shocks, and also obtained clues to the effects of resonance scattering, X-ray emitting gas, and non-radiative shocks. In particular, it has been confirmed that the resonance scattering in the regions viewed edge-on is much higher than in the other regions. The X-ray and FUV luminosities of the Cygnus Loop calculated from a revised distance have been compared with those of the Vela SNR. In the Cygnus Loop, the X-ray and \ion{O}{6} luminosities relative to those of the other lower ionization FUV lines are higher than the Vela SNR. This suggests that the Cygnus Loop contains a much larger fraction of shocks in the early evolutionary stages than the Vela SNR.

\acknowledgements{}
SPEAR/FIMS is a joint project of the Korea Astronomy and Space Science Institute, the Korea Advanced Institute of Science and Technology, and the University of California at Berkeley, funded by the Korea MOST and NASA grant NAG5-5355. I.-J. Kim was supported by a National Research Foundation of Korea grant funded by the Korean government. We thank the referee for comments that improved the manuscript.

\end{document}